\documentclass[a4paper,preprint,showpacs,superscriptaddress]{revtex4}
\usepackage[version=3]{mhchem}
\usepackage{float,fancyhdr,amsmath,graphicx,subfig,epsfig,color,setspace,url}
\usepackage{appendix}
\begin{document}

\title{Dissociative Electron Attachment to Polyatomic Molecules - III : Ammonia }
\author{N. Bhargava Ram}
\email[]{nbhargavaram@tifr.res.in}
\affiliation{Tata Institute of Fundamental Research, Mumbai 400005, India}

\author{E. Krishnakumar}
\email[]{ekkumar@tifr.res.in}
\affiliation{Tata Institute of Fundamental Research,  Mumbai 400005, India}

\begin{abstract}

In this paper, we discuss the dissociative electron attachment process in Ammonia. Kinetic energy and angular distributions of \ce{H-} and \ce{NH2^{-}} fragment anions across the two well known resonances at 5.5 eV and 10.5 eV are reported. The angular distributions show deviation of axial recoil approximation akin to that observed in water.  
\end{abstract}

\maketitle

\section{Introduction}

Following the measurements on water and hydrogen sulphide (molecules with three atoms), we proceed to study the dissociation dynamics in Ammonia, the next smallest polyatomic molecule constituting four atoms. Ammonia is known to play a major role in the synthesis of bigger organic molecules in the interstellar medium, amino acids. The formation of amino acids by the irradiation of ultraviolet light and high energy electron beams on mixtures of ice containing ammonia and other small molecules has been demonstrated experimentally \cite{c5greenberg,c5bernstein,c5holtom}. Ammonia is a constituent of the atmospheres of many planets and comets. The measurements on Ammonia also assume importance in the context of engineering applications such as plasma reactors for waste treatment or plasma surface treatment.

\ce{NH3} has a bent pyramidal equilibrium geometry in ground state with electronic configuration \ce{(1a1)^{2} (2a1)^{2} (1e)^{4} (3a1)^{2} -> ^{1}A1} in \ce{C_{3v}} geometry. Ammonia is known to have electron attachment resonance peaks at 5.5 eV and 10.5 eV. Table \ref{tab5.1} lists the various dissociation channels possible with their thresholds and corresponding anion resonance states based on Wigner-Witmer correlation rules. Earlier reports showed that the \ce{NH3^{-*}} resonance at these energies dissociates to give \ce{H-} and \ce{NH2^{-}} fragments and measured their cross sections and kinetic energies \cite{c5sharpdowell,c5compton,c5stricklett,c5tronc}. Based on energetics and correlations with vacuum ultra violet absorption spectrum/electron energy loss experiment data, there have been speculations on the possible symmetry states accessed and the dissociation mechanisms \cite{c5sharpdowell}. It is necessary to measure the angular distribution of the fragments as they are direct signatures of the symmetry of the resonance and the dissociation dynamics. So far, there exists only one angular distribution measurement reported by Tronc et al. \cite{c5tronc} of the \ce{H-} ions at 5.7 eV. Whereas, there are no measurements reported from the second resonance at 10.5 eV. In this chapter, we present the angular and kinetic energy distributions of the \ce{H-} and \ce{NH2^{-}} fragments at the two resonances obtained using the velocity map imaging technique. 

\begin{table}
\caption{Various dissociation channels with threshold and corresponding symmetries of \ce{NH3^{-*}} based on Wigner-Witmer rules}
\begin{center}
\begin{tabular}{ccc}
\hline
\\
Dissociation channel & Threshold & Symmetry \\
\\
\hline
\\
\ce{H- + NH2(^{2}B1)} &3.76 eV & \ce{A1} \\
\\
\ce{H- + NH2^{*}(^{2}A1)} &5.03 eV & \ce{A1} or E \\
\\
\ce{H + NH2^{-}(^{1}A1)} &3.30 eV & \ce{A1} or E \\
\\
\ce{H + NH2^{*-}(^{1}A1)} &5.78-EA(\ce{NH2^{*}}) & \ce{A1} or E \\
\\
\ce{H- + H + NH}($^{2}\Sigma^{-}$) &7.57 eV & \ce{A2} \\
\\
\ce{H + H + NH-}($^{2}\Pi$) &7.94 eV & E \\
\\
\hline
\end{tabular}
\end{center}
\label{tab5.1}
\end{table}

\section{Earlier work}
One of the earliest works on DEA to Ammonia was reported by Sharp and Dowell \cite{c5sharpdowell}. They measured the cross sections for total negative ion yield as a function of energy and isotope effects in \ce{NH3} and \ce{ND3}. Two peaks were seen at 5.65 and 10.5 eV producing both \ce{H-} and \ce{NH2^{-}}. By measuring the kinetic energies using the retarding potential difference method, they deduced that the DEA products at the first resonance of 5.65 eV are in electronic ground state. At the 10.5 eV resonance, they speculate that the \ce{NH2} fragment produced with \ce{H-} is in the electronically excited state. 

Compton et al.\cite{c5compton} using the \ce{SF6} scavenger method and trapped electron technique studied the electron impact excitation of ammonia and detected the \ce{^{1}A2^{"}} peak (lowest optically excited state of ammonia) at about 6 eV - the parent state of the DEA resonance at 5.65 eV. Another peak is observed between 10 and 11 eV. They also determined the absolute cross section values for the negative ion production due to dissociative electron attachment in \ce{NH3} and \ce{ND3}. 

A comprehensive study of the DEA process at the first resonance was done by Stricklett and Burrow \cite{c5stricklett} concluding that the \ce{NH3^{-*}} resonance state decays to produce fragment ions via a predissociation mechanism similar to that in neutral excited state of Ammonia \cite{c5douglas}. They looked at the vibrational spacing in the cation ground state, lowest excited Rydberg state of the neutral and the anion to find that the spacing are approximately close and attributed to the out of plane $\nu_{2}$ vibration mode. Similar comparison was done for \ce{ND3} as well confirming the $\nu_{2}$ mode excitation. This close similarity suggested that the temporary negative ion dissociates in the same way as the lowest excited Rydberg state by predissociation. Further, the presence of two channels i.e. \ce{H- (^{1}S) + NH2(^{2}B1)} and \ce{H(^{2}S)+NH2^{-}(^{1}A1)} with different symmetry requires an avoided crossing between the surfaces. They also concluded that the planar dissociation of the temporary anion produces \ce{H-} whereas non planar dissociation gives rise to \ce{NH2^{-}}. 

Tronc et al. \ce{tronc} measured the differential cross section with respect to electron energy in addition to  kinetic energy and angular distribution of the \ce{H-} and \ce{NH2^{-}} fragments across the resonance at 5.6 eV. The vibrational structure seen in the ion yield curves of the two fragments were in agreement with those of Stricklett and Burrow \cite{c5stricklett}. The vibrational progression in the ion yield curve of both the anions were attributed to the $\nu_{2}$ out of the plane vibrational mode and predissociation mechanism and charge transfer between the two \ce{A^{$\prime$}} resonance states leading to \ce{H- + NH2} and \ce{H + NH2-} products at infinite separation in distorted geometry. They also observed that there is no valence predissociating state that could be detected for the neutral state but a broad shape resonance ($\sigma^{*}$ \ce{NH3^{-}}) has been observed at around 7 eV in the vibrational excitation of the \ce{NH3} electronic ground state. This resonance state overlaps the \ce{^{2}A2^{"}} \ce{NH3^{-}}  Feshbach resonance and can predissociate its quasibound out of the plane vibrational levels $n\nu_{2}$. Angular distribution of the \ce{H-} ions at 5.7 eV was also reported by them.

\section{Angular distributions for $C_{3v}$ point group}

Azria et al. \cite{c5azria} showed that the theory of O'Malley and Taylor \cite{c5omalleytaylor} calculating the angular distribution of the fragment anions from DEA to diatomic molecules could be extended to polyatomic molecules. In paper-I on water, the angular distribution curves for various symmetries under the \ce{C_{2v}} point group were calculated within the axial recoil approximation. The expressions obtained from these calculations were used to fit the angular distribution data of Water and Hydrogen Sulphide and infer the symmetry of the molecular negative ion that dissociated. We now proceed to calculate the angular distribution curves for various symmetries under \ce{C_{3v}} point group to be applied to the case of DEA to Ammonia. A \ce{C_{3v}} molecule has 6 symmetry operations which turn the molecule into itself, namely, Identity (I), Rotations by $\pm 120^{\circ}$ about the \ce{C3} axis and reflections in the 3 mirror planes formed by each of the NH bonds and the \ce{C3} axis ($\sigma_{\nu1}$,$\sigma_{\nu2}$ and $\sigma_{\nu3}$). Reduced to three conjugacy classes based on similarities in the operations, the symmetry states (or the 'representations') in this point group are \ce{A1}, \ce{A2} and E, where \ce{A1} and \ce{A2} are one dimensional representations and E is a two-dimensional representation. The character table for the \ce{C_{3v}} point group and the basis functions written in terms of spherical harmonics transforming irreducibly under this group \cite{c5wohlecke,c5koster} are in given Table \ref{tab5.2}.

\begin{table}
\caption{Character table for \ce{C_{3v}} point group and basis functions}
\begin{center}
\begin{tabular}{ccccc}
\hline
\\
 & I & \ce{2C2} & 3$\sigma_{\nu}$ & Basis functions \\
\\
\hline
\\
\ce{A1} & 1 & 1 & 1 & $Y_{l}^{0}; l=0,1,2,3$ \\
&&&& $Y_{3}^{3} - Y_{3}^{-3}$ \\
\\
\ce{A2} & 1 & 1 & -1 & $Y_{3}^{3} + Y_{3}^{-3}$ \\
\\
E & 2 & -1 & 0 & $(Y_{l}^{-1},-Y_{l}^{1}); l=1,2,3$ \\
(2 dim representation) &&&& $(Y_{l}^{2},Y_{l}^{-2}); l=2,3$ \\
\\
\hline
\end{tabular}
\end{center}
\label{tab5.2}
\end{table}

\begin{figure}[!ht]
\centering
\subfloat[\ce{A1 -> A1}]{\includegraphics[width=0.4\columnwidth]{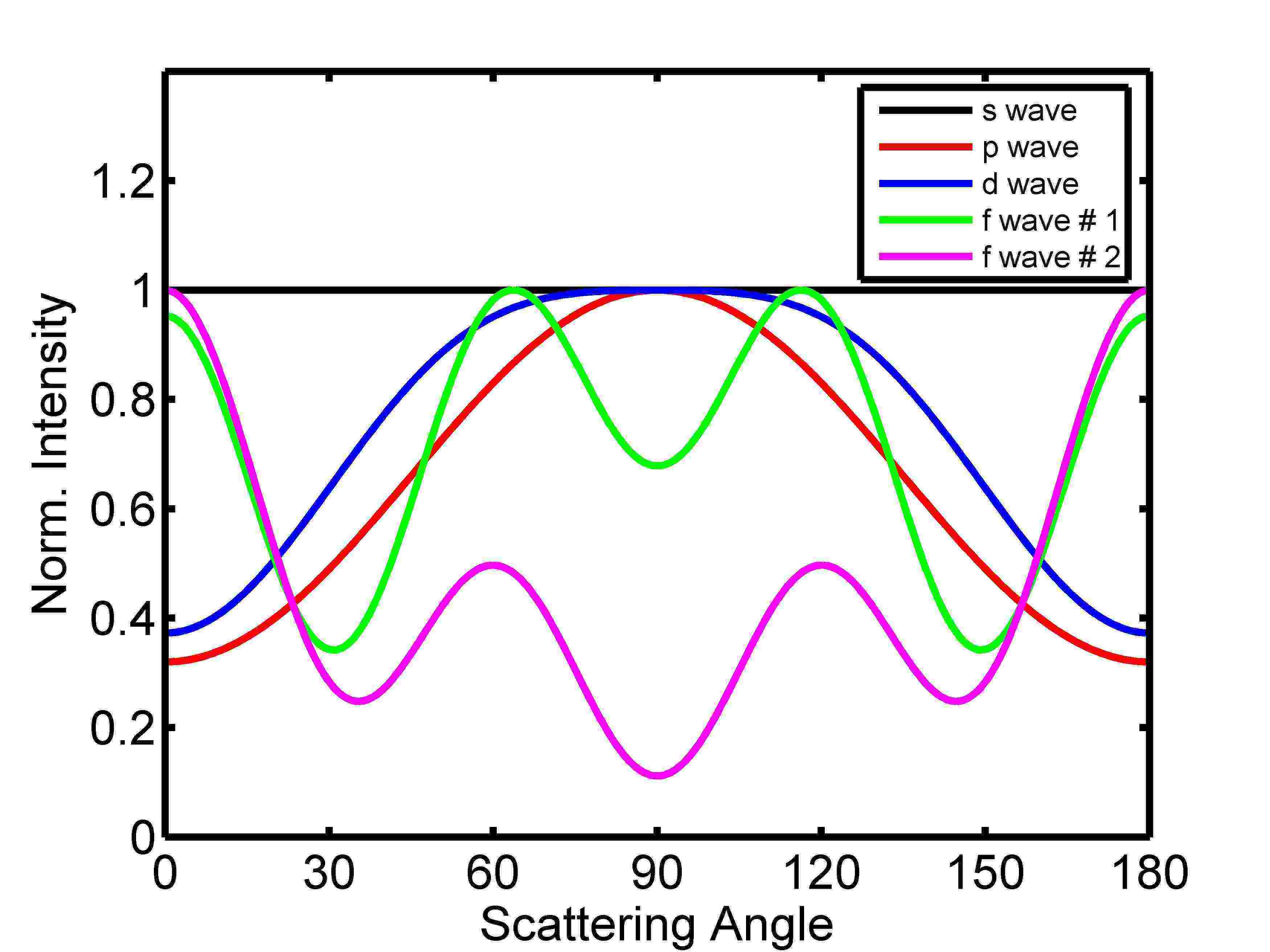}}
\subfloat[\ce{A1 -> A2}]{\includegraphics[width=0.4\columnwidth]{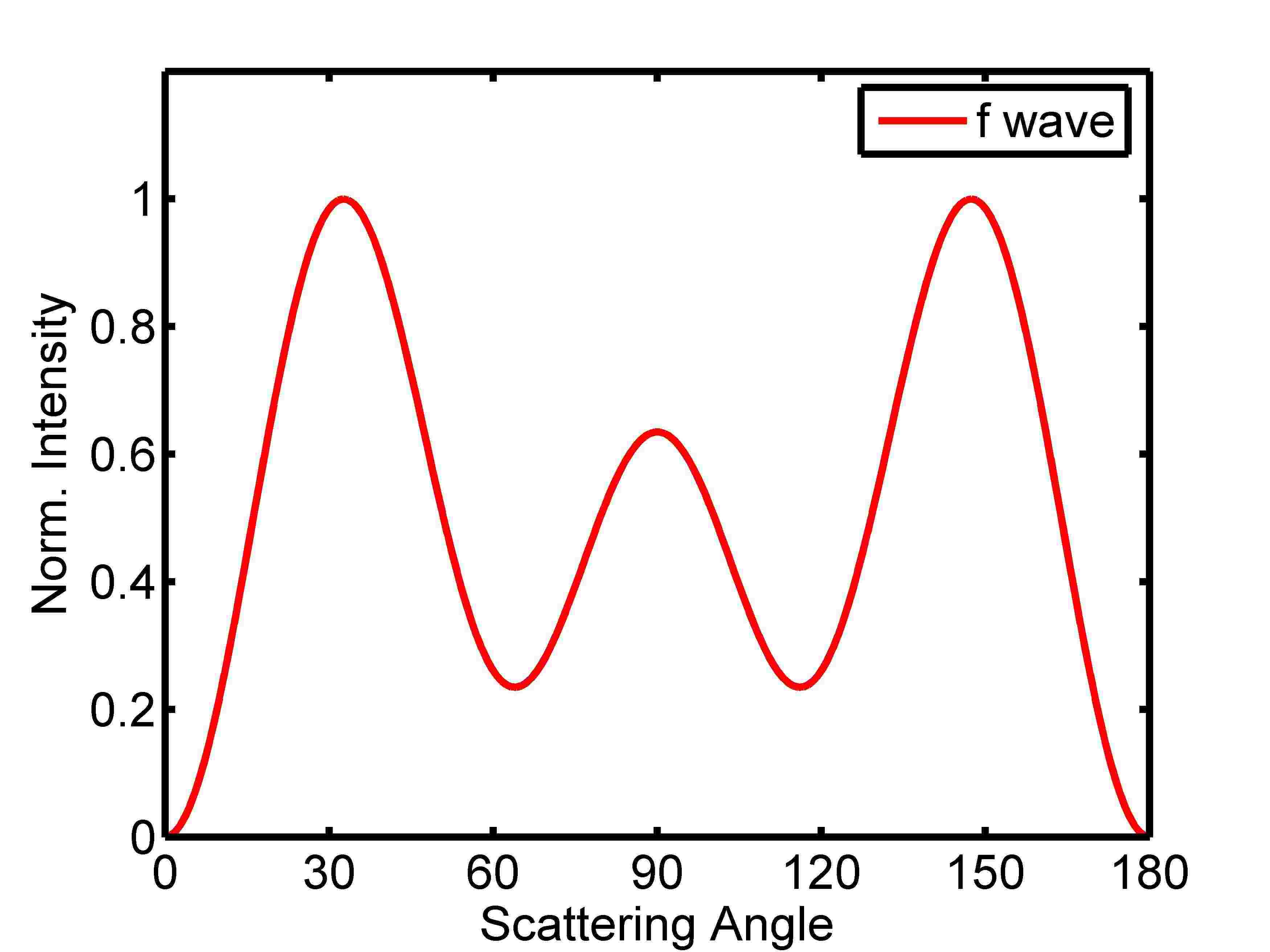}}\\
\subfloat[\ce{A1 -> E}]{\includegraphics[width=0.4\columnwidth]{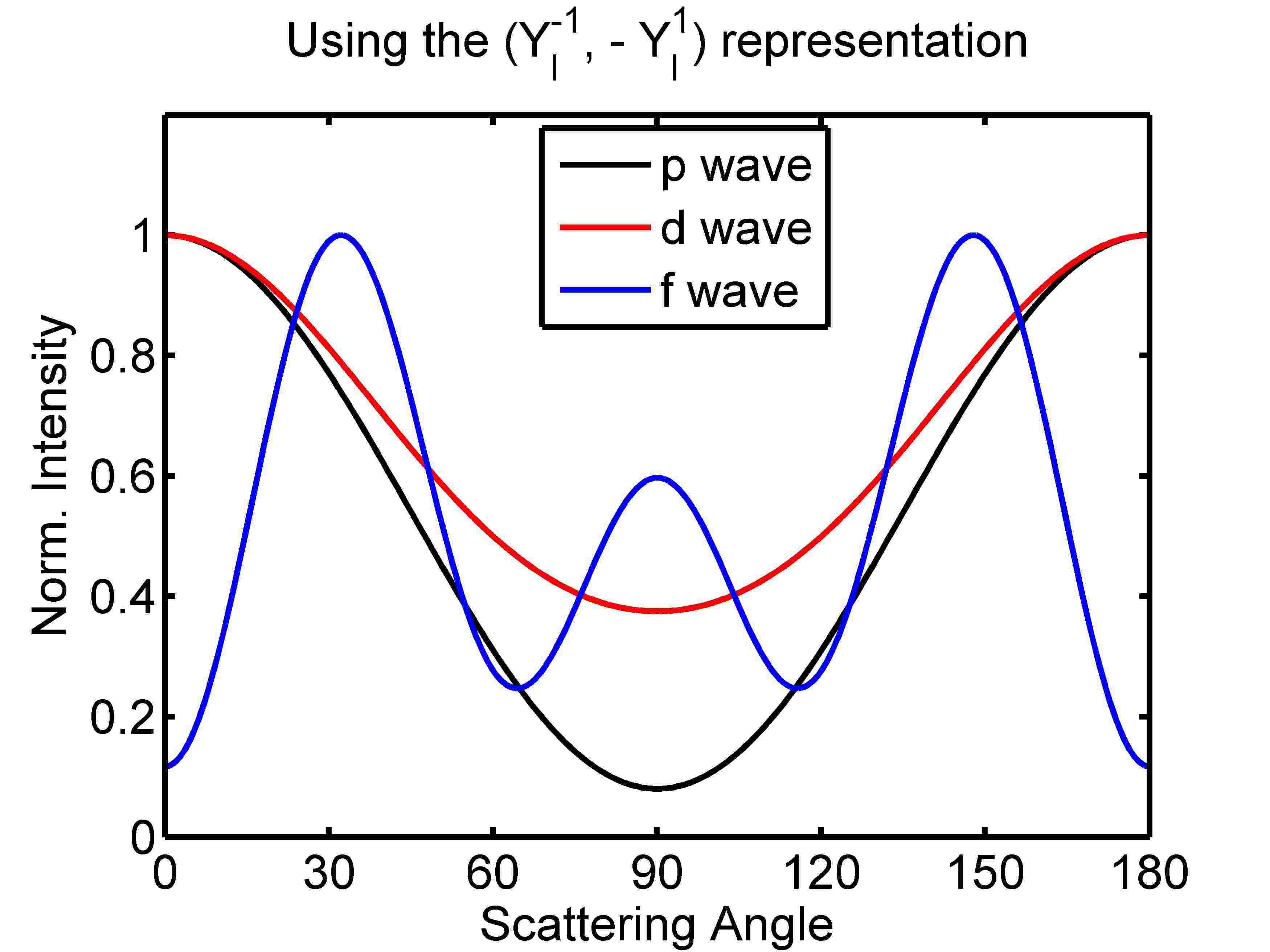}}
\subfloat[\ce{A1 -> E}]{\includegraphics[width=0.4\columnwidth]{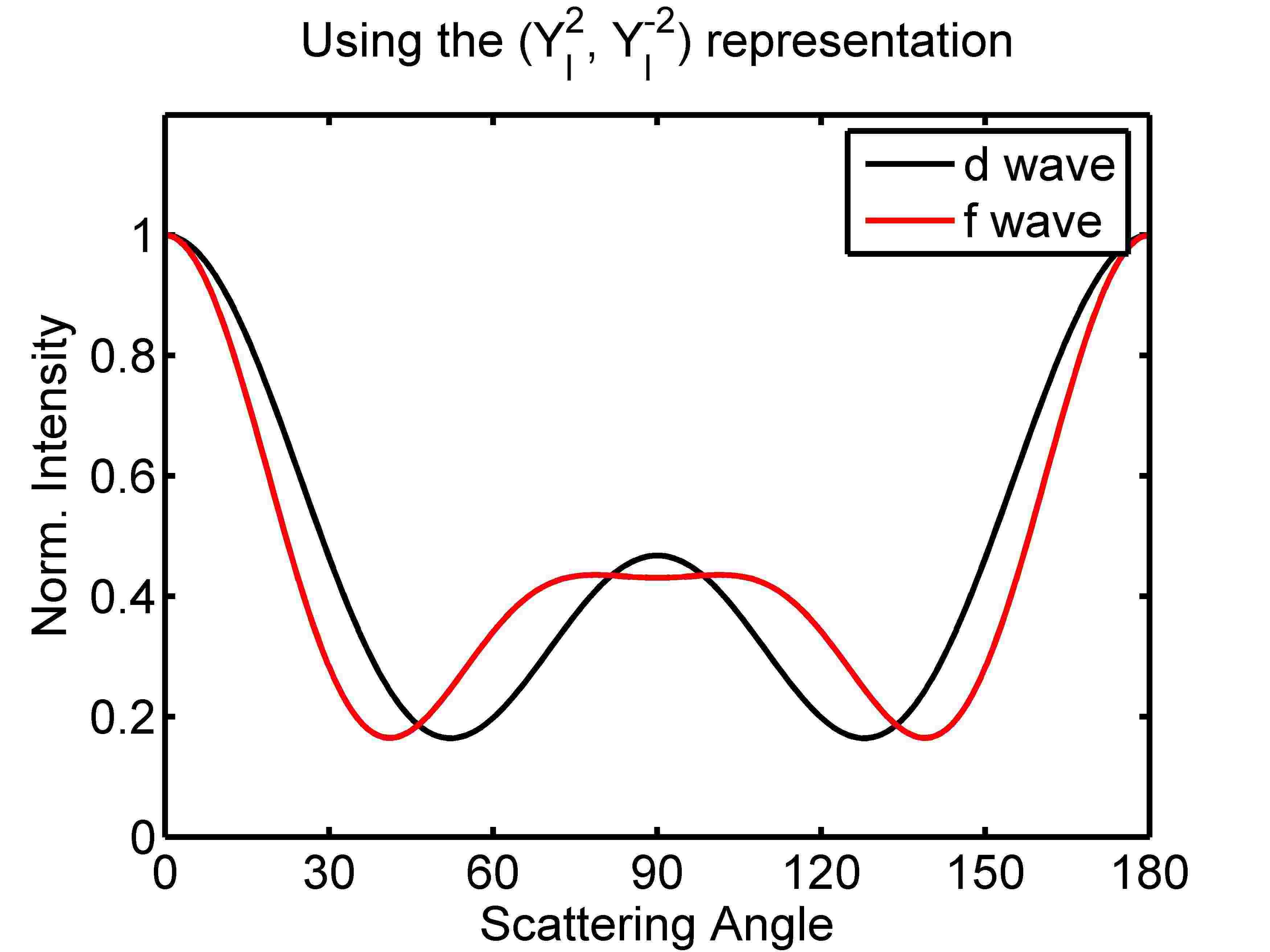}}
\caption{Angular distribution plots for the \ce{A1}, \ce{A2} and E states for various allowed partial waves. The curves are normalized to their maximum value.}
\label{fig5.1}
\end{figure}

The transition amplitude i.e. $<Negative Ion State|Partial Wave|Initial Neutral State>$ is calculated and consequently, this quantity is squared and integrated over the azimuthal angles (due to $\phi$ symmetry in the scattering process) to obtain the scattering intensity I($\theta$). Considering electron attachment to Ammonia molecule in its electronic ground state (which is \ce{A1}), we calculate amplitudes for formation of a negative ion of \ce{A1}, \ce{A2} and E respectively taking various partial waves. The partial wave (representing the electron beam) in lab frame is transformed to the dissociation frame by the Euler angles $(\phi,\theta,0)$ and the electronic states \ce{A1}, \ce{A2} and E (defined in molecular frame w.r.t \ce{C3} axis) are transformed to the dissociation axis by Euler angles $(0,\beta,0)$. Angle $\beta$ is the angle between the \ce{C3} axis and one of the dissociating N-H bonds. In ground state equilibrium geometry, this angle is $68.2^{\circ}$. We have used this value to deduce the final expression for the angular distribution. The expressions for transition amplitudes for various negative ion states are calculated similar to that in paper I and the final scattering intensities are deduced. Figure \ref{fig5.1} shows angular distribution curves for \ce{H-} ions dissociating from the ammonia anion of various symmetries assuming ground state equilibrium geometry. Interference due to mixing of two or more partial waves is taken in consideration while fitting the angular data. The scattering intensity expressions obtained under \ce{C_{3v}} group for partial waves upto $l=3$ are listed in Appendix A given at the end of this paper. 

\section{Results and Discussion}

\begin{figure}[!h]
\centering
\includegraphics[width=0.6\columnwidth]{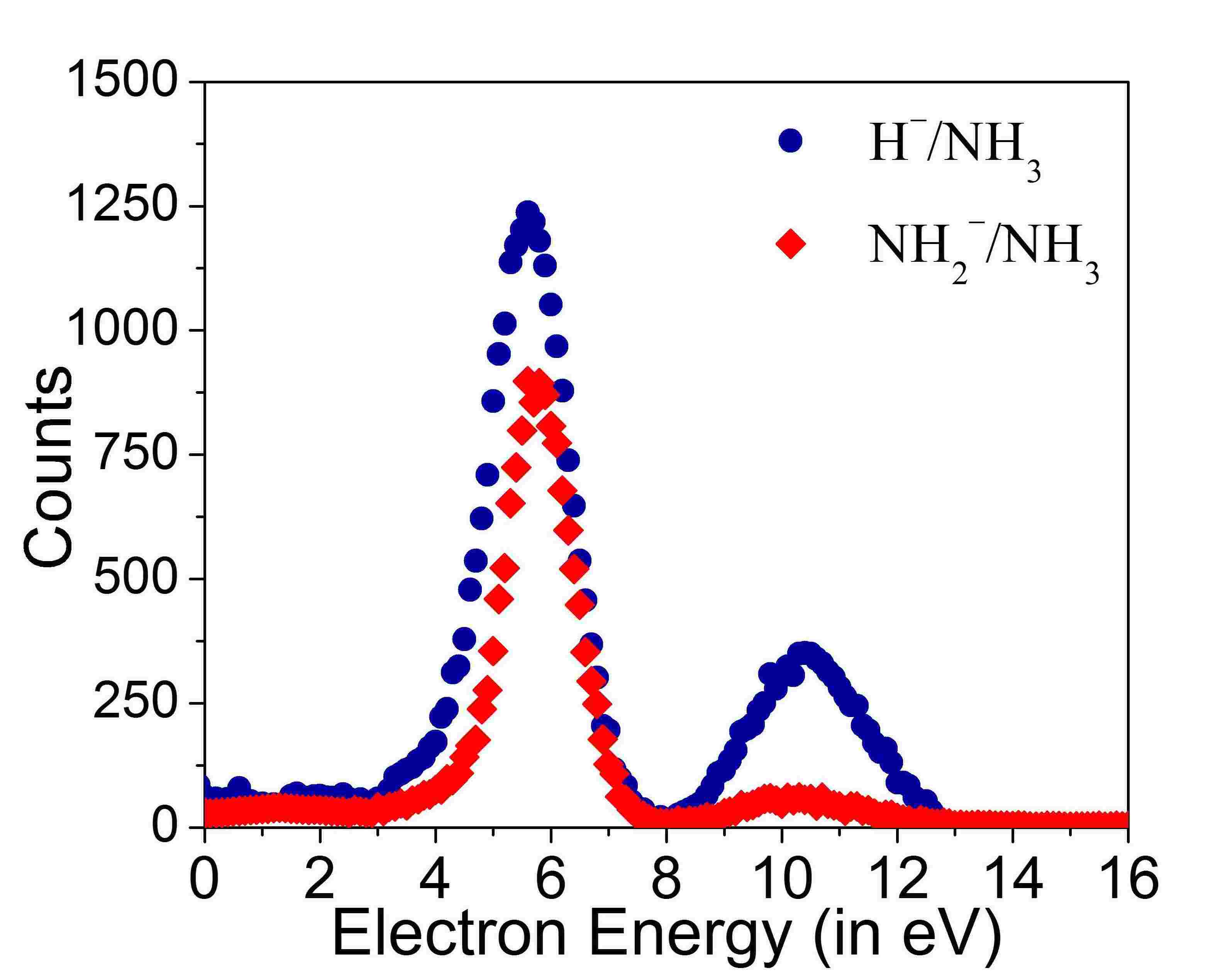}
\caption{ Ion yield curve for \ce{H-} and \ce{NH2^{-}}. (Not to scale)}
\label{fig5.2}
\end{figure}

\begin{figure}[!htbp]
\centering
\subfloat[\ce{H-} at 4.5 eV]{\includegraphics[width=0.3\columnwidth]{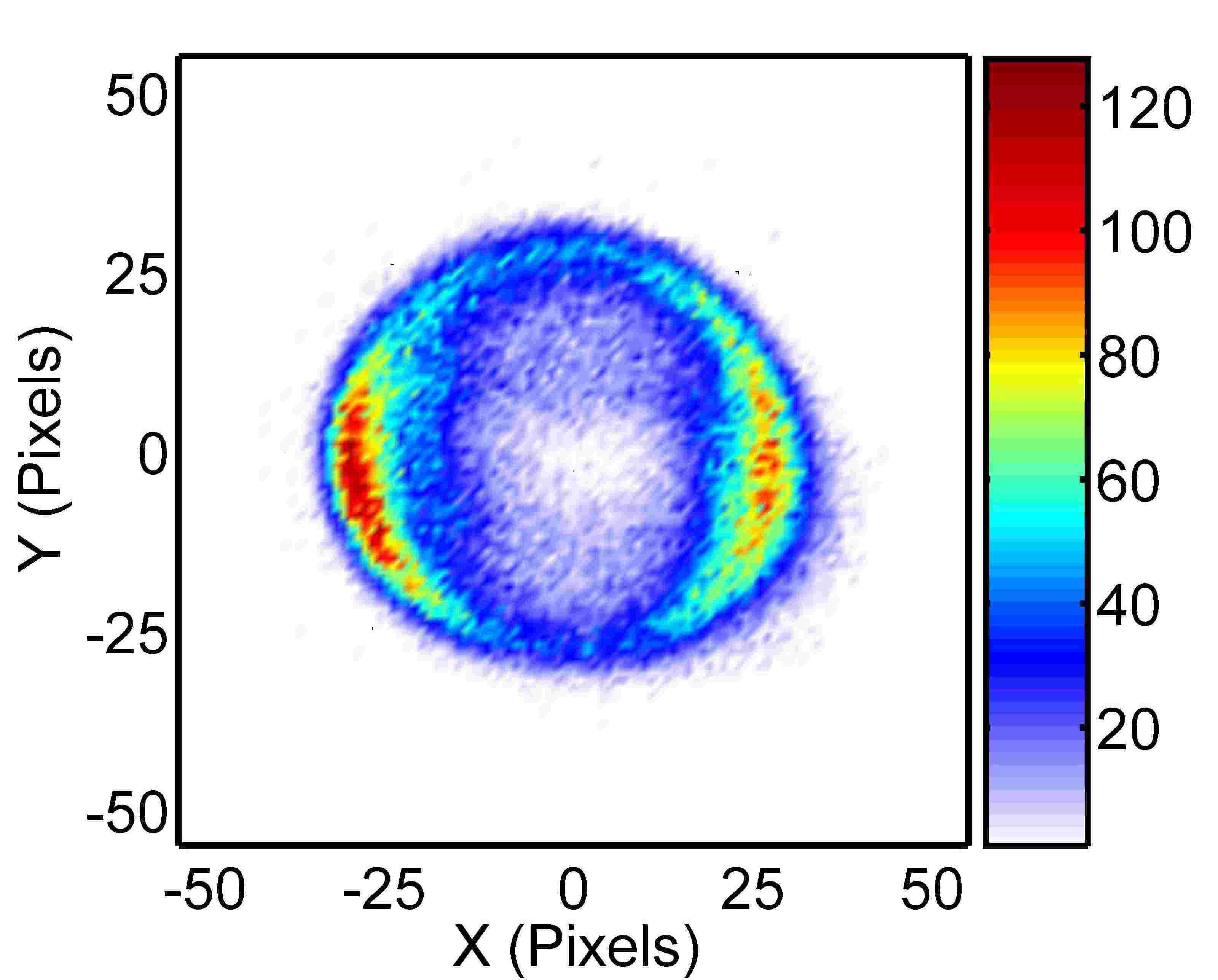}}
\subfloat[\ce{H-} at 5.5 eV]{\includegraphics[width=0.3\columnwidth]{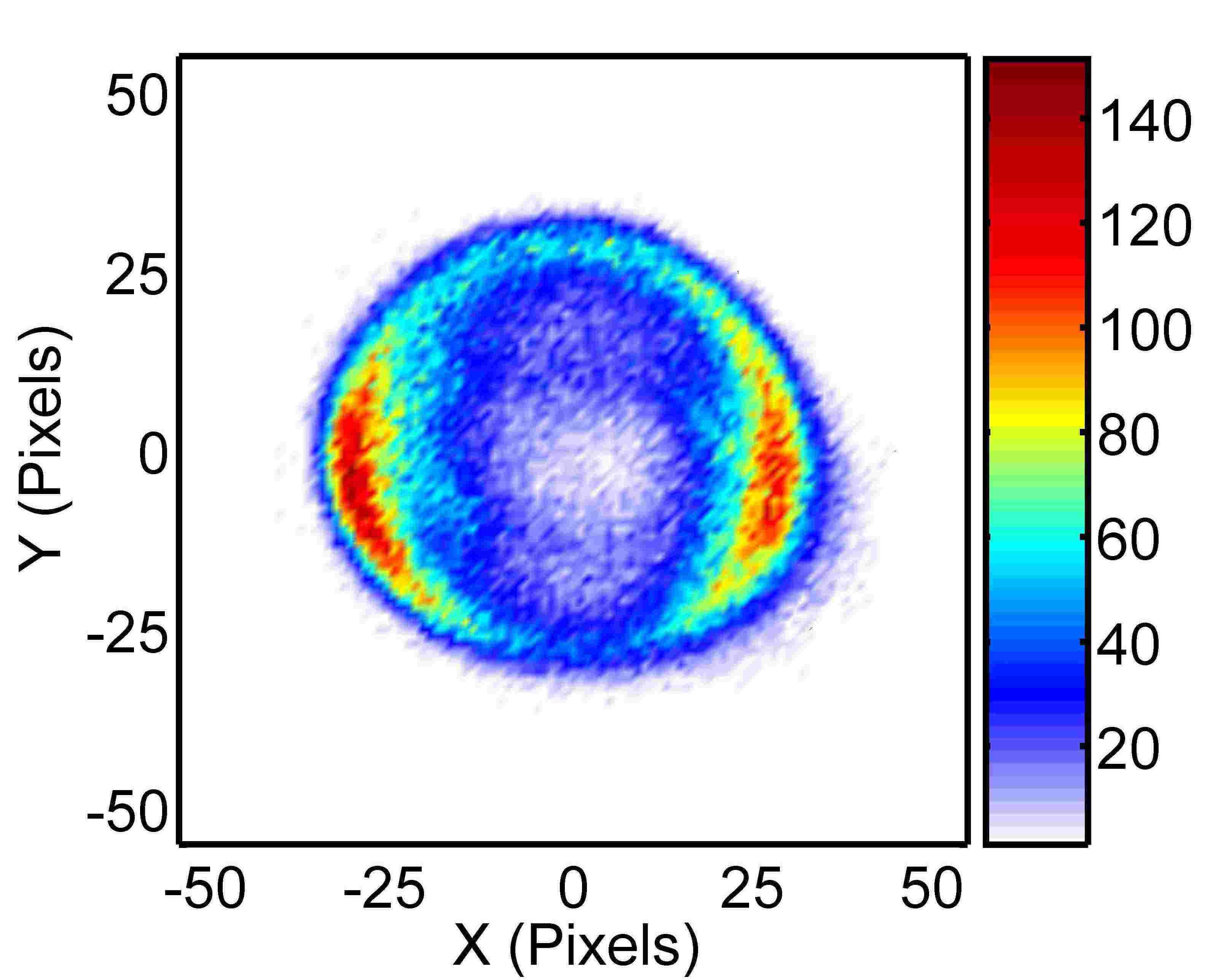}}
\subfloat[\ce{H-} at 6.5 eV]{\includegraphics[width=0.3\columnwidth]{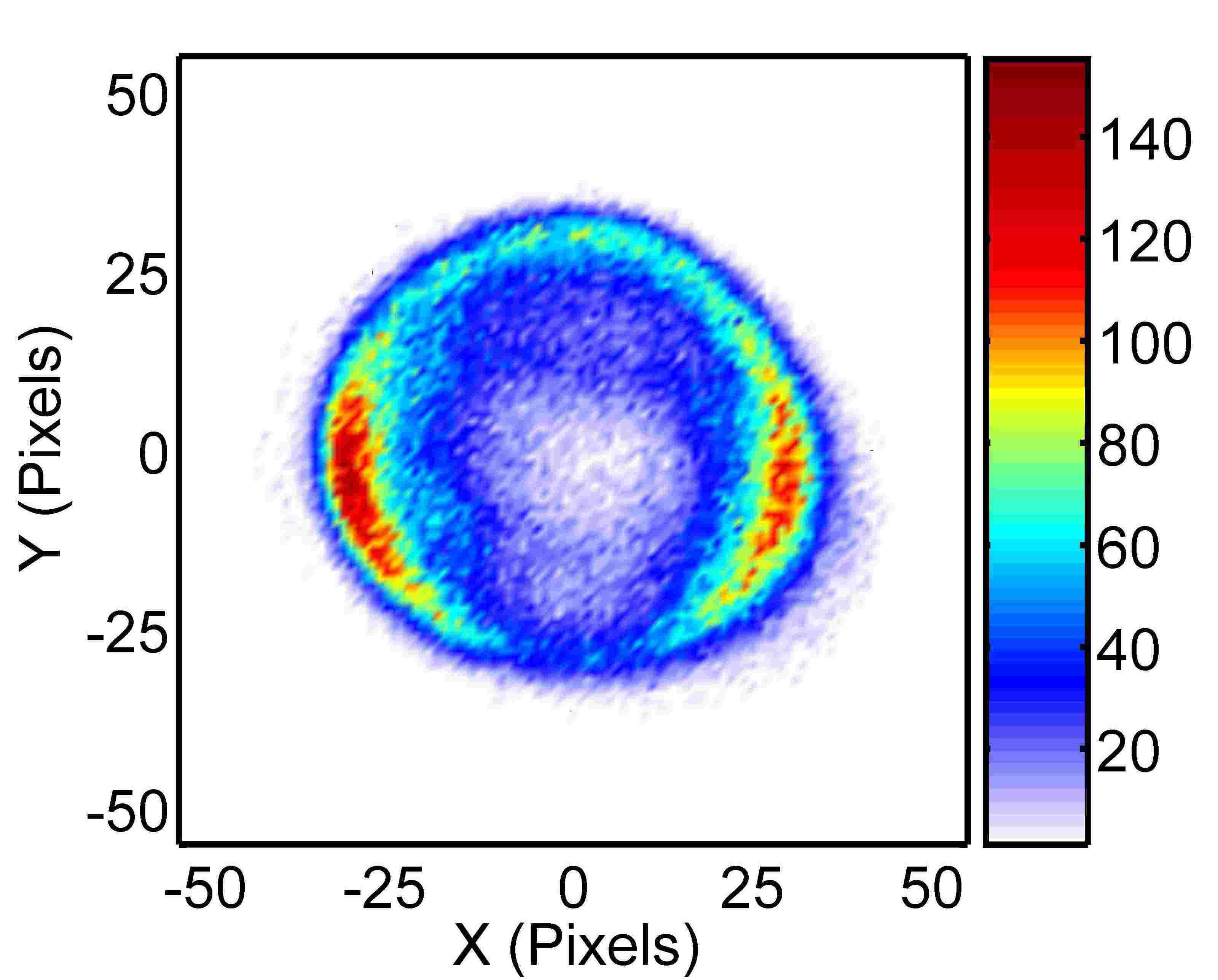}}\\
\subfloat[\ce{H-} at 9.5 eV]{\includegraphics[width=0.3\columnwidth]{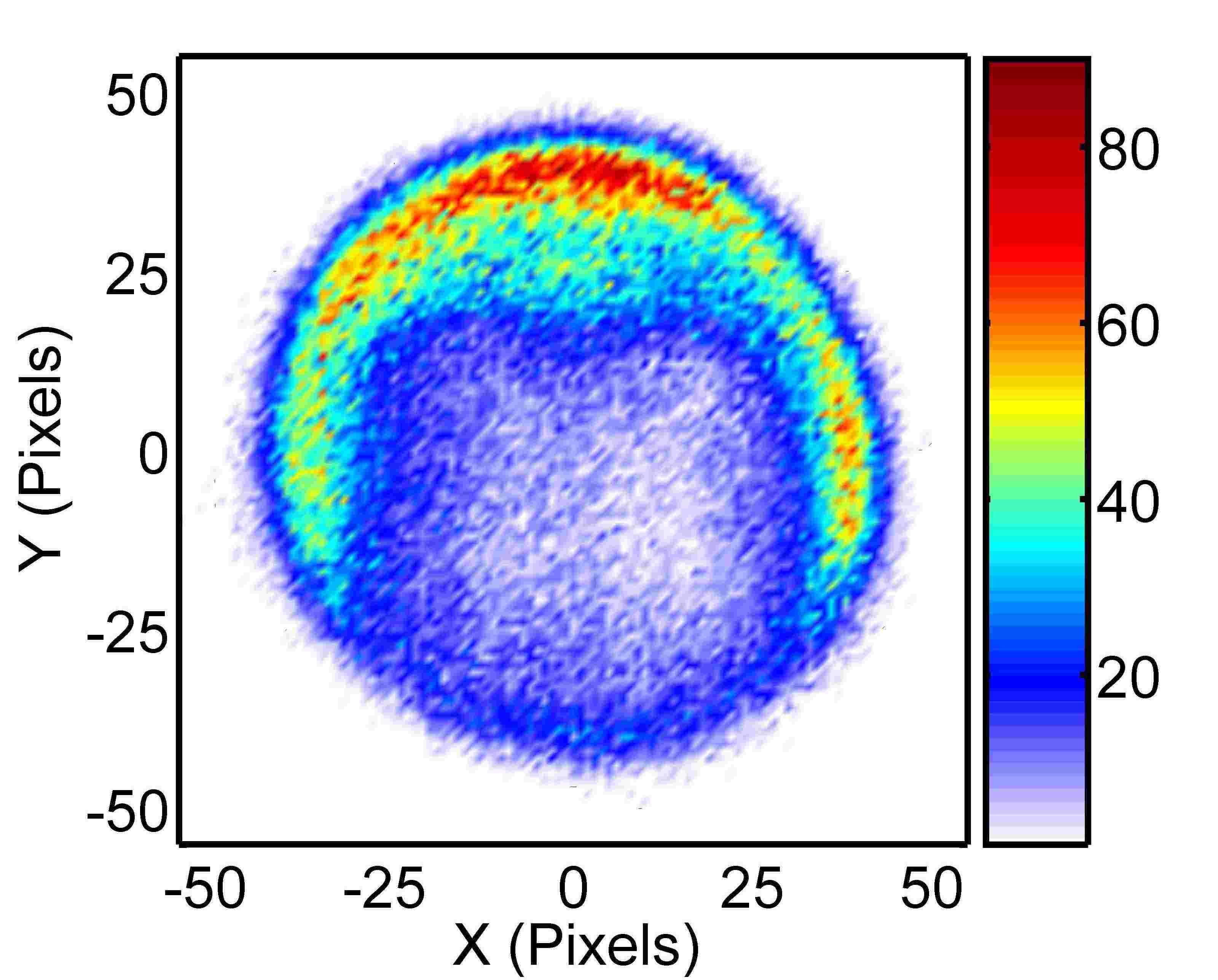}}
\subfloat[\ce{H-} at 10.5 eV]{\includegraphics[width=0.3\columnwidth]{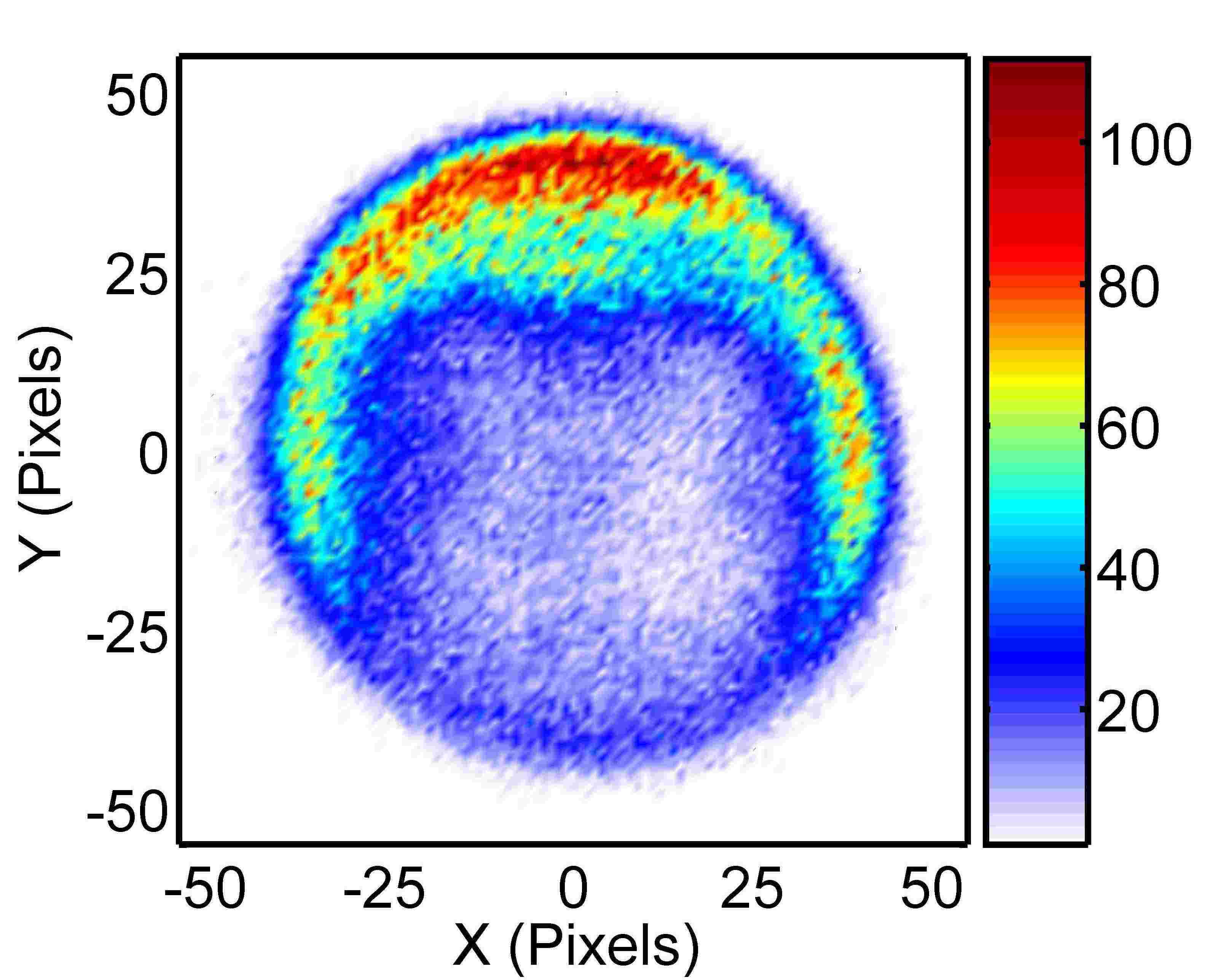}}
\subfloat[\ce{H-} at 11.5 eV]{\includegraphics[width=0.3\columnwidth]{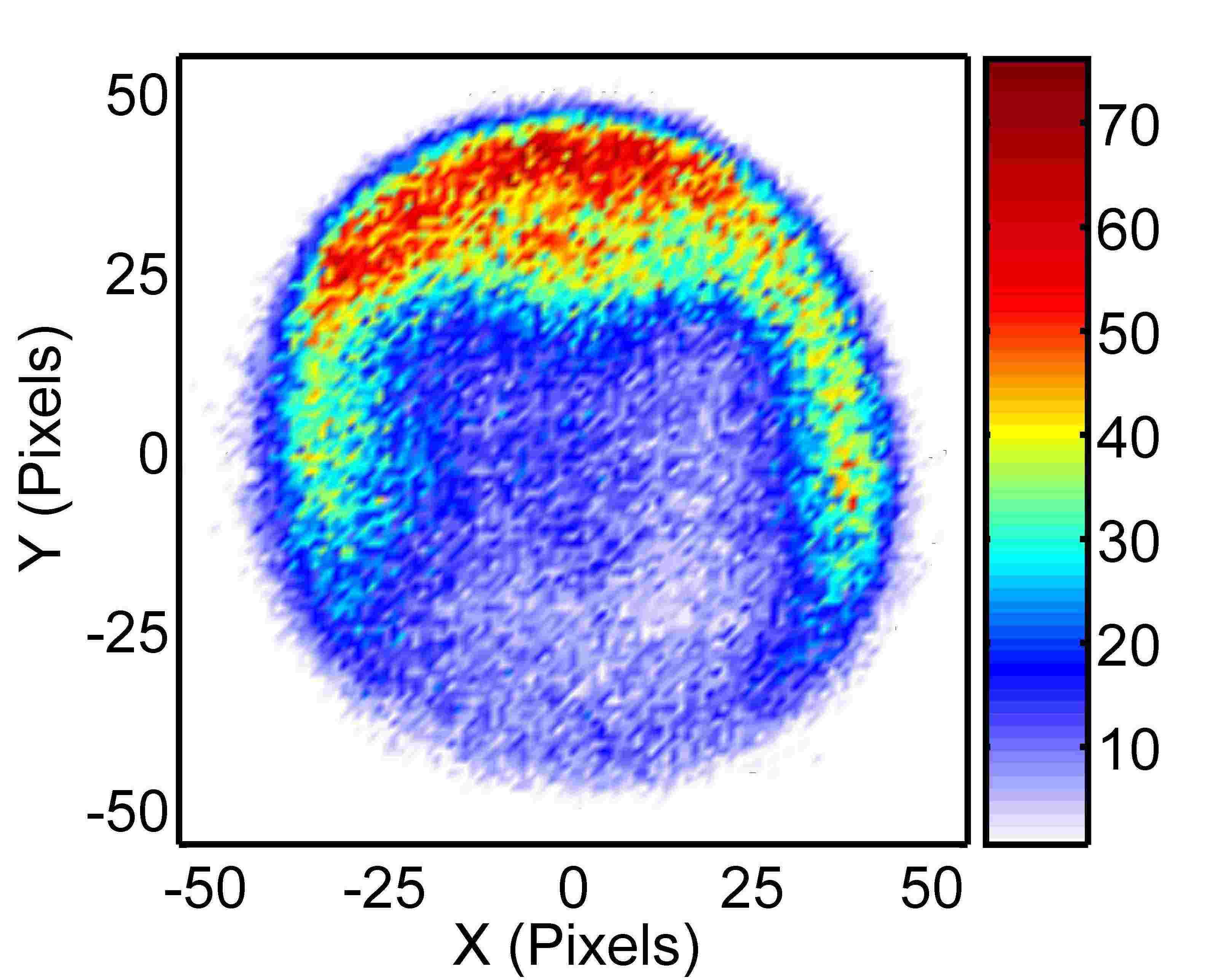}}
\caption{Velocity Images of \ce{H-} ions from DEA to \ce{NH3}. The electron beam is from top to bottom in every image.}
\label{fig5.3}
\end{figure}

\begin{figure}[!htbp]
\centering
\subfloat[\ce{NH2^{-}} at 4.5 eV]{\includegraphics[width=0.3\columnwidth]{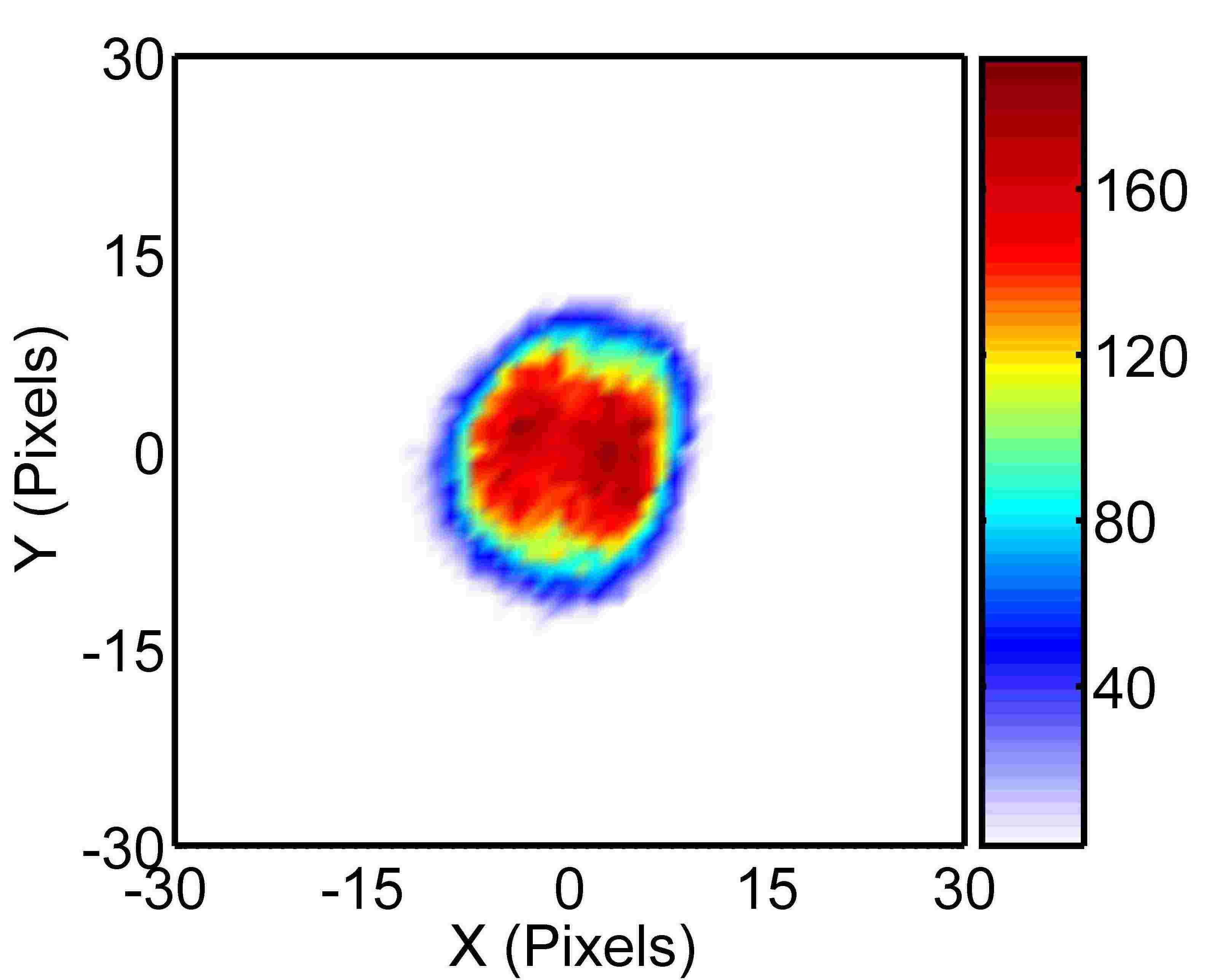}}
\subfloat[\ce{NH2^{-}} at 5.5 eV]{\includegraphics[width=0.3\columnwidth]{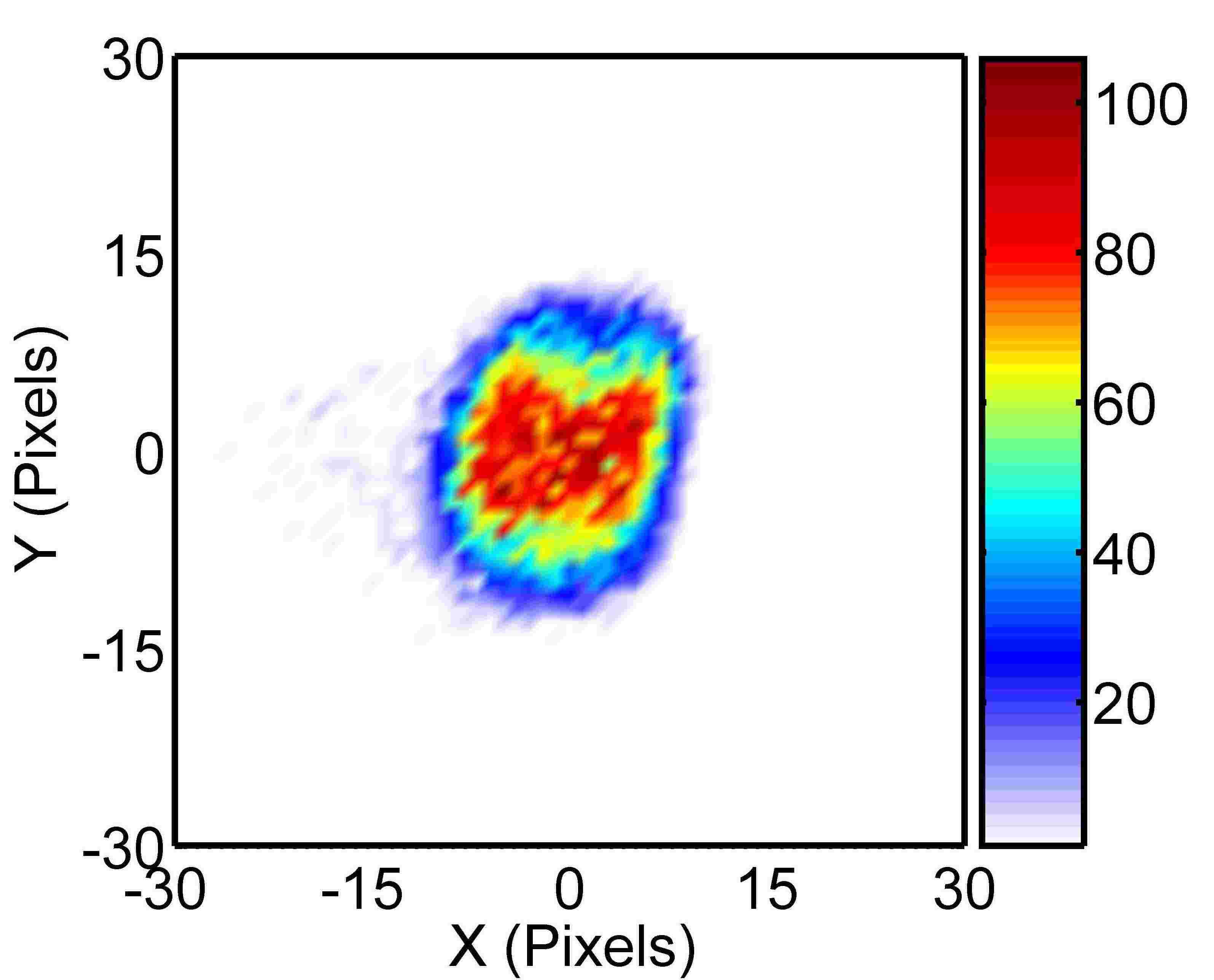}}
\subfloat[\ce{NH2^{-}} at 6.5 eV]{\includegraphics[width=0.3\columnwidth]{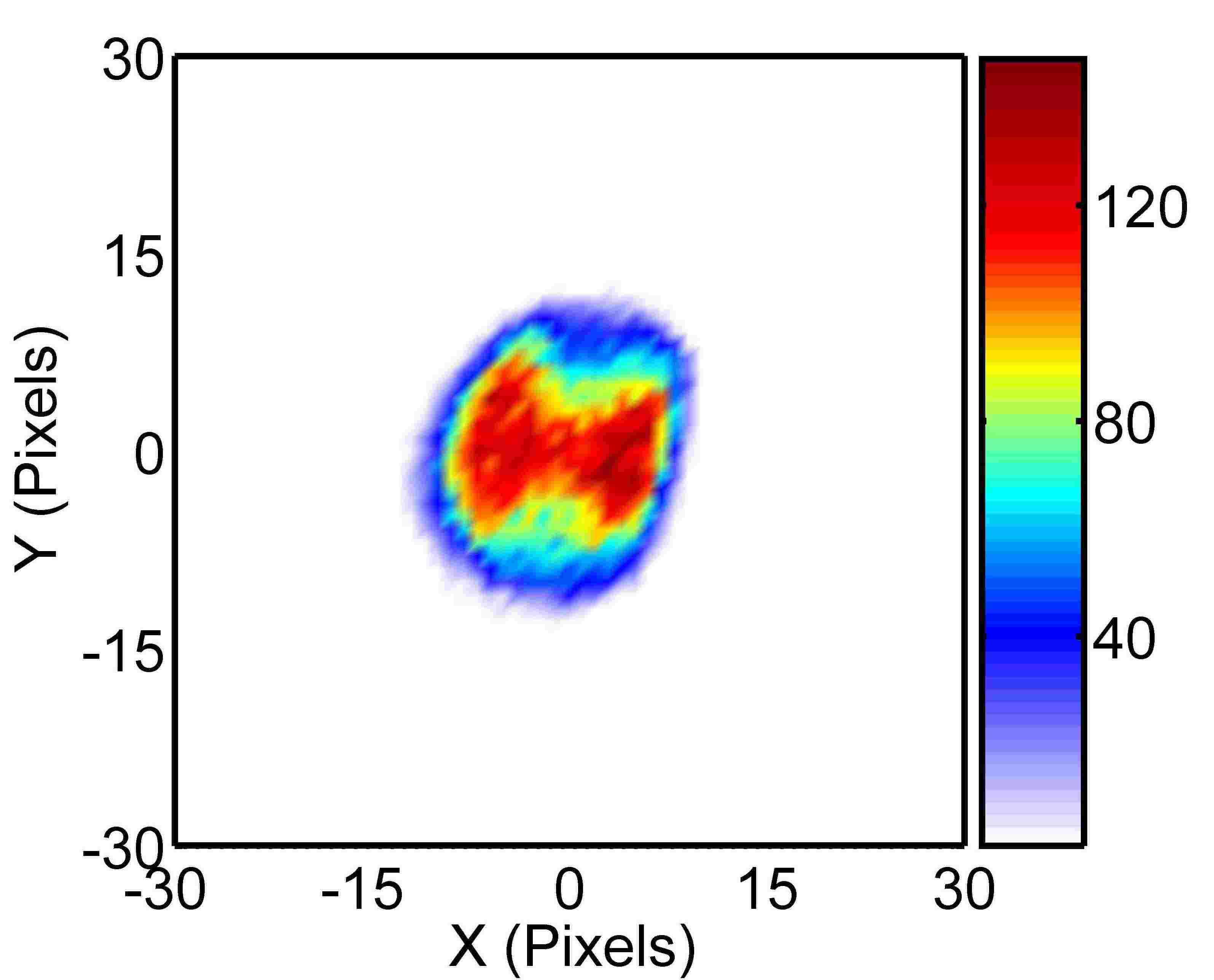}}\\
\subfloat[\ce{NH2^{-}} at 9.5 eV]{\includegraphics[width=0.3\columnwidth]{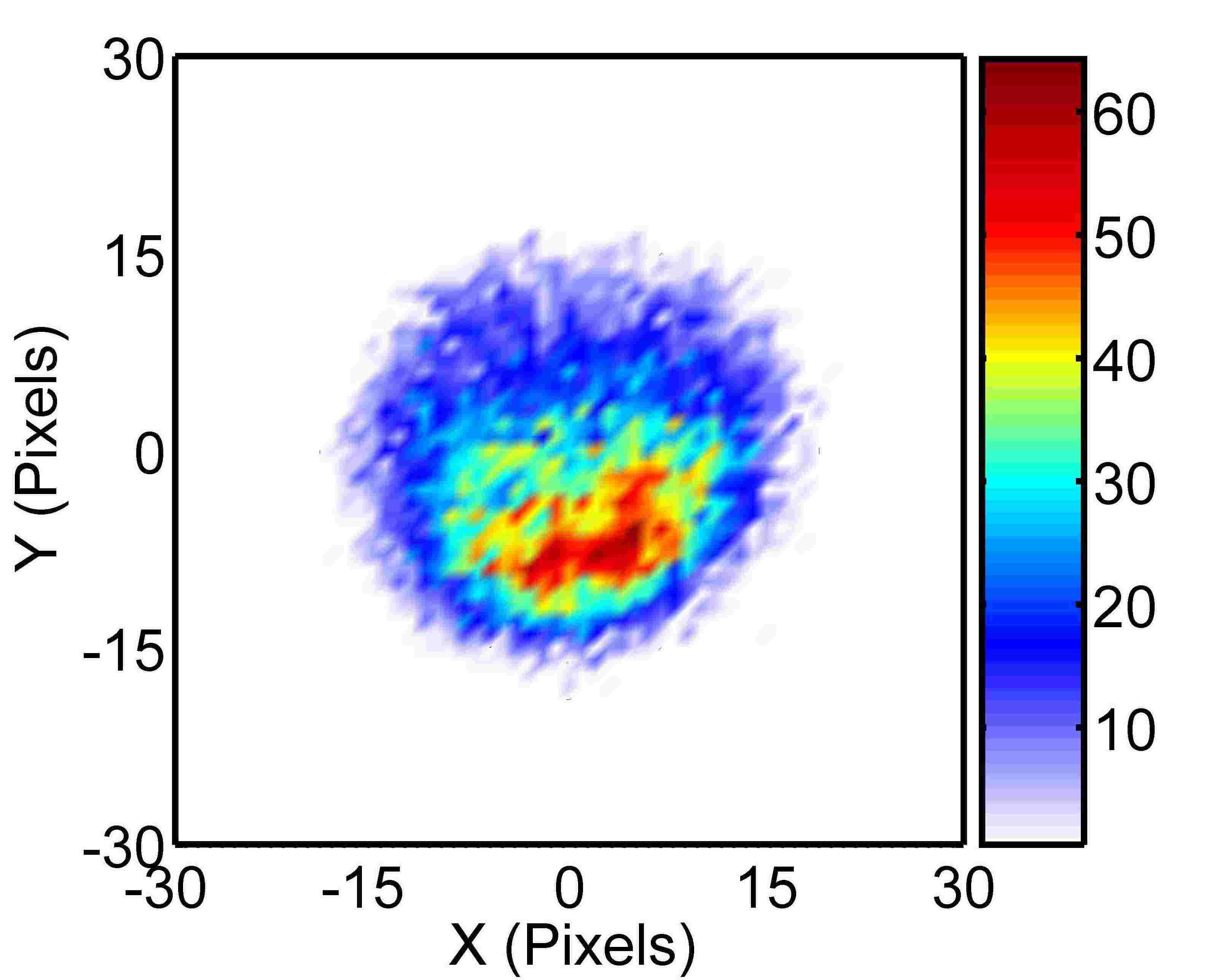}}
\subfloat[\ce{NH2^{-}} at 10.5 eV]{\includegraphics[width=0.3\columnwidth]{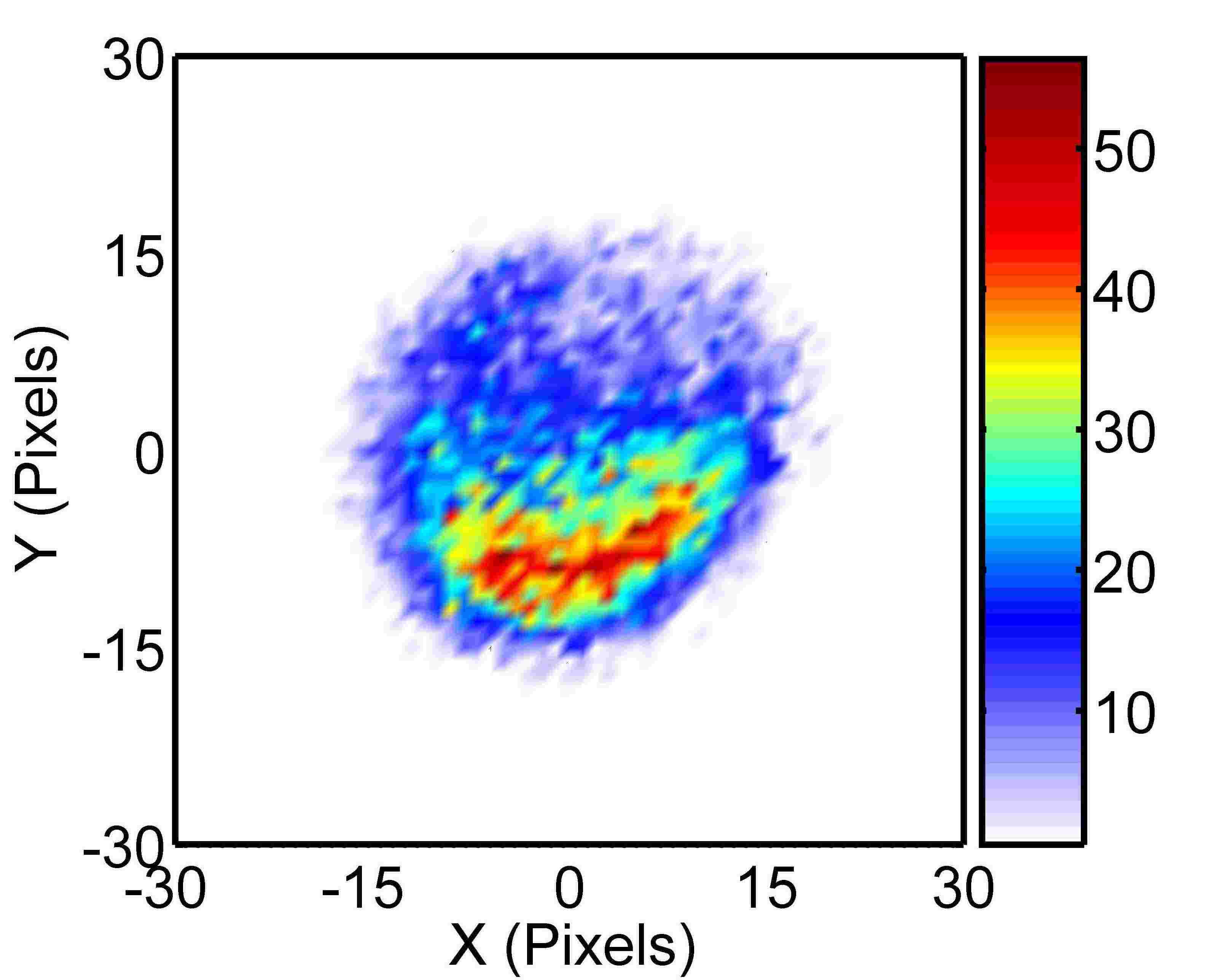}}
\subfloat[\ce{NH2^{-}} at 11.5 eV]{\includegraphics[width=0.3\columnwidth]{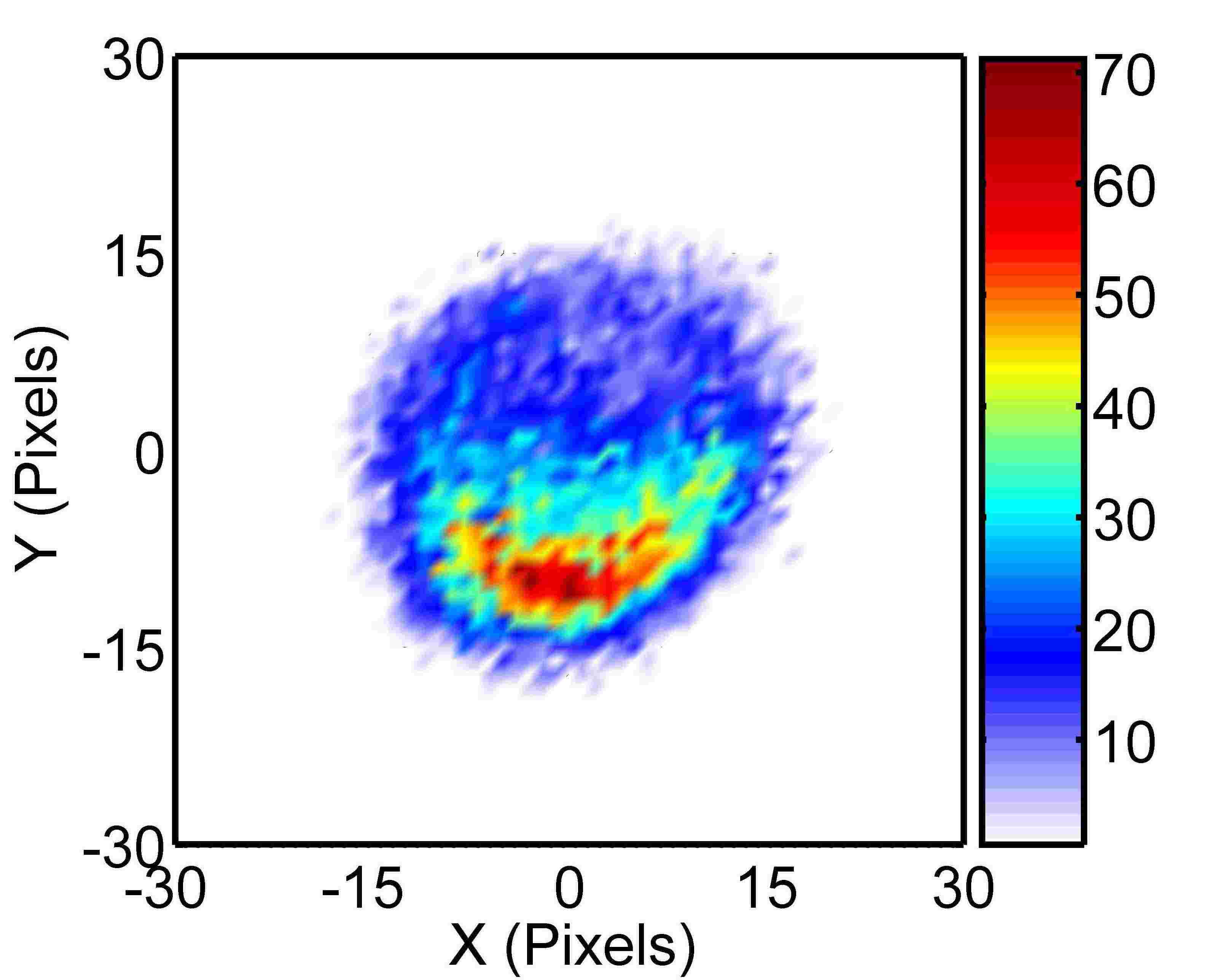}}
\caption{Velocity Images of \ce{NH2^{-}} ions from DEA to \ce{NH3}. The electron beam is from top to bottom in every image.}
\label{fig5.4}
\end{figure}

The ion yield curve of \ce{H-} and \ce{NH2^{-}} produced from DEA to \ce{NH3} (Figure \ref{fig5.2}) shows two resonance structures peaking at 5.5 eV and 10.5 eV. This is consistent with earlier measurements. Our spectrometer cannot separate \ce{NH-} and \ce{NH2^{-}} masses. However, the presence of \ce{NH-} as a product has been ruled out by Sharp and Dowell \cite{c5sharpdowell} from their high resolution measurements. The angular and kinetic energy distributions are discussed in the following sections.  The velocity images of \ce{H-} and \ce{NH2^{-}} across the two resonances are shown in Figures \ref{fig5.3} and \ref{fig5.4} respectively.

\subsection{First resonance process peaking at 5.5 eV}

\begin{figure}[!htbp]
\centering
\subfloat[]{\includegraphics[width=0.4\columnwidth]{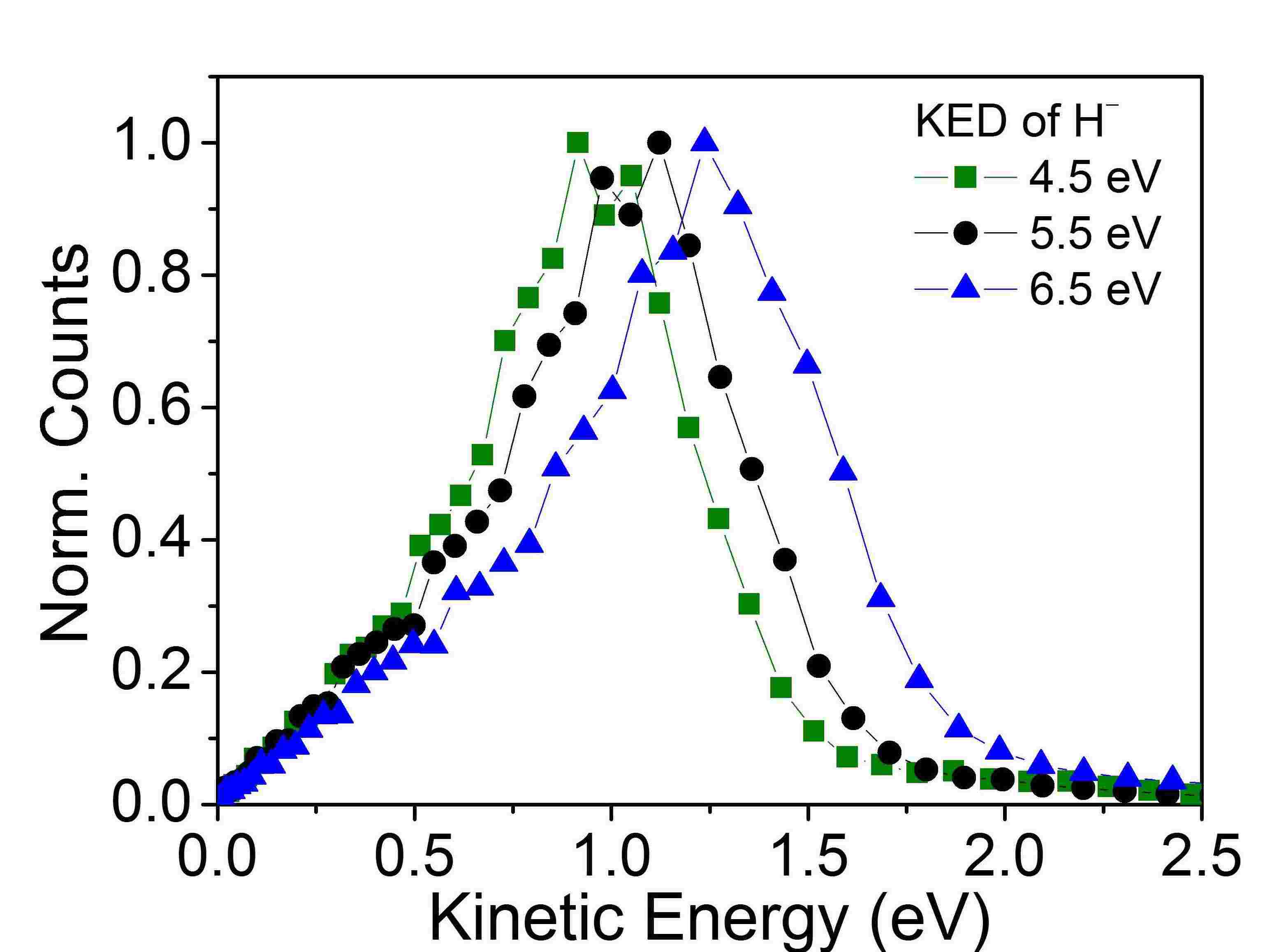}}
\subfloat[]{\includegraphics[width=0.4\columnwidth]{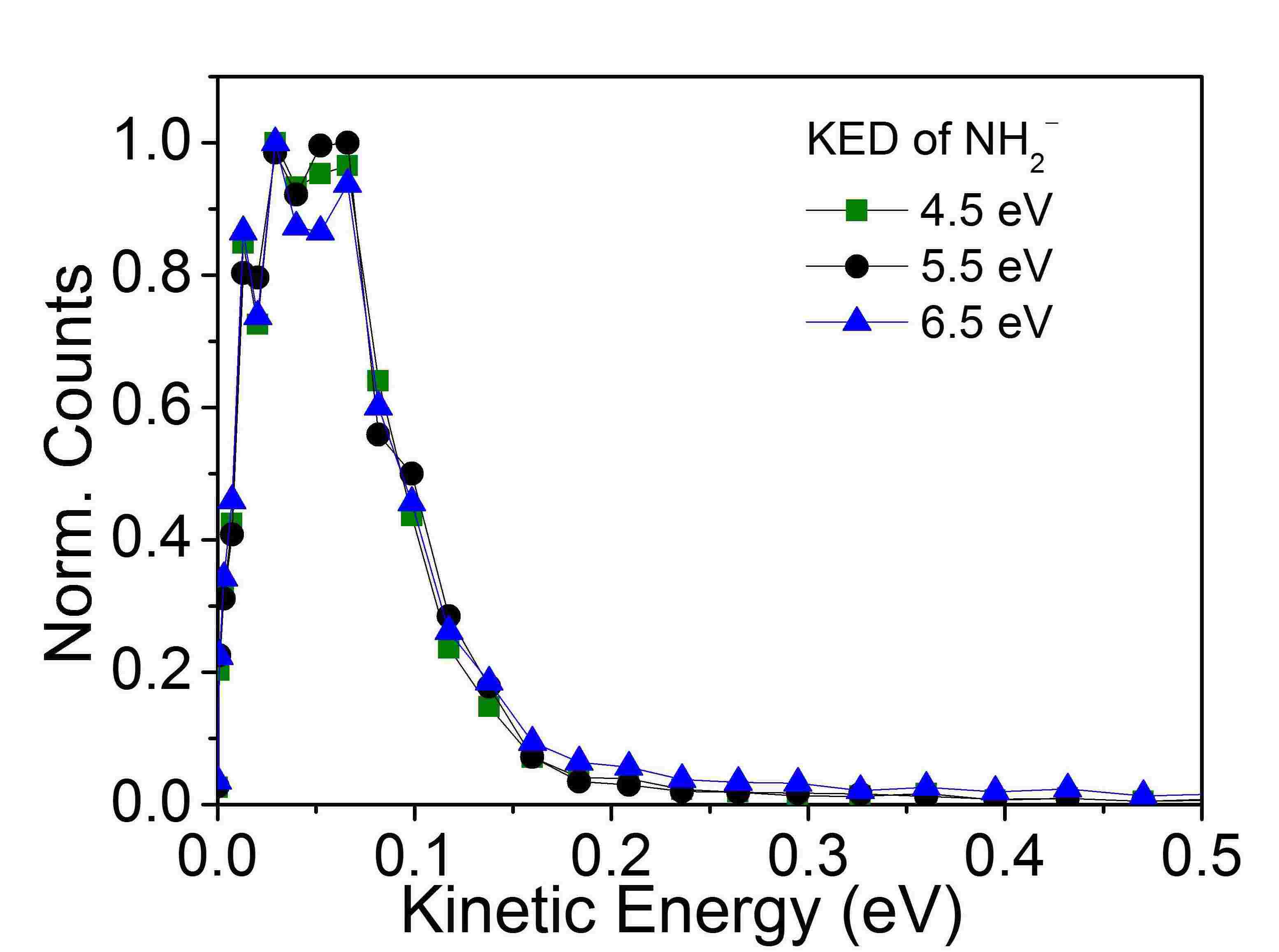}}
\caption{KE distribution of (a) \ce{H-} and (b) \ce{NH2^{-}} ions across the first resonance at 4.5, 5.5 and 6.5 eV}
\label{fig5.5}
\end{figure}

Figures \ref{fig5.3}(a), (b) and (c) show the velocity images of \ce{H-} ions at electron energies 4.5 eV, 5.5 eV and 6.5 eV. The electron beam is from top to bottom in every image. The \ce{H-} ions are mostly scattered perpendicular to the electron beam direction with finite intensity in the forward and backward angles. The backward angles appear more intense than the forward part. Figure \ref{fig5.5}(a) shows the kinetic energy distribution of the \ce{H-} ions across the first resonance peak at electron energies 4.5, 5.5 and 6.5 eV. At the peak of the resonance i.e. 5.5 eV, the KE ranges from 0 to about 1.8 eV and points to \ce{H- (^{1}S) + NH2(^{2}B1)} channel with threshold 3.76 eV. To elaborate, about 4.51 eV is needed to break the H-\ce{NH2} bond and the electron affinity of H atom is 0.75 eV. Hence, when the incident electron energy is 5.5 eV, the excess energy is about 1.7 eV and the maximum kinetic energy of \ce{H-} is estimated to be 16/17th of 1.7 eV i.e. 1.6 eV. This is close to the observed value of 1.8 eV. The broad distribution indicates internal excitation (vibrational and rotational) of the \ce{NH2} fragment. Owing to the poorer energy resolution ($\sim$0.5 eV) of the electron beam, we are unable to see distinct rings in the velocity map image of \ce{H-} corresponding to the vibrational excitation of the \ce{NH2} fragment. Presence of the \ce{H- + NH2^{*}(^{2}B1)} channel with threshold of 5.03 eV is ruled out as the kinetic energy of \ce{H-} ions in that case would be about 0.5 eV and we do not see any specific \ce{H-} ion group/lump appearing in the velocity image corresponding to 0.5 eV. From Figure \ref{fig5.5}(a), we see that with increase in electron energy, the maximum KE of \ce{H-} increases and the distribution becomes broader. Thus, the excess energy is taken away as translational energy of the fragments and also in the internal excitation of the \ce{NH2} fragment.

Figure \ref{fig5.4}(a), (b) and (c) show the velocity images of the \ce{NH2^{-}} ions at 4.5 eV, 5.5 eV and 6.5 eV respectively. Inspite of the small size of the image, it is discernible that the \ce{NH2^{-}} ions are ejected perpendicular to the electron beam similar to \ce{H-} ions. The maximum kinetic energy of the \ce{NH2^{-}} ions is found to be lower than 0.2 eV (see Figure \ref{fig5.5}(b)). The possible dissociation channel is understood to be \ce{H(^{2}S) + NH2^{-}(^{1}A1)} with appearance energy of 3.30 eV. Based on Wigner-Witmer rules, this channel correlates to an $A_{1}$ resonance (see Table \ref{tab5.1}).  Estimating the maximum kinetic energy of \ce{NH2^{-}} ions from this channel, we get about 0.13 eV which is 1/17th of the excess energy above threshold of 3.30 eV. This is close to our observed value of about 0.2 eV. This confirms the presence of \ce{H(^{2}S) + NH2^{-}(^{1}A1)} channel and is consistent with the findings of Sharp and Dowell \cite{c5sharpdowell}. Sharp and Dowell also speculated the possibility of \ce{H + NH2^{-*}} channel, where the amino anion is in the first electronically excited state. So far, there are no measurements available in literature ascertaining the electron affinity of the \ce{NH2} radical in the first electronic excited state. Hence, the appearance energy for the latter is not known accurately. Assuming the appearance energy of such a channel to be close to 5 eV (taking into account the excitation energy of the first excited state and the electron affinity), the kinetic energy of \ce{NH2^{-*}} would be close to zero eV. Considering that the stability of the \ce{NH2^{-*}} is not known and that \ce{NH2^{-}} in ground electronic state (\ce{^{1}A1}) explains the observed KE and angular distribution, we conclude that the channel producing the heavier anion fragment is \ce{H + NH2^{-}(^{1}A1)}. However, it is very much desirable that experiments with increased energy resolution and theoretical calculations on topology of potential energy surfaces address the question of \ce{NH2^{-*}} ion (in the first excited state).

\begin{figure}[!htbp]
\centering
\subfloat[]{\includegraphics[width=0.45\columnwidth]{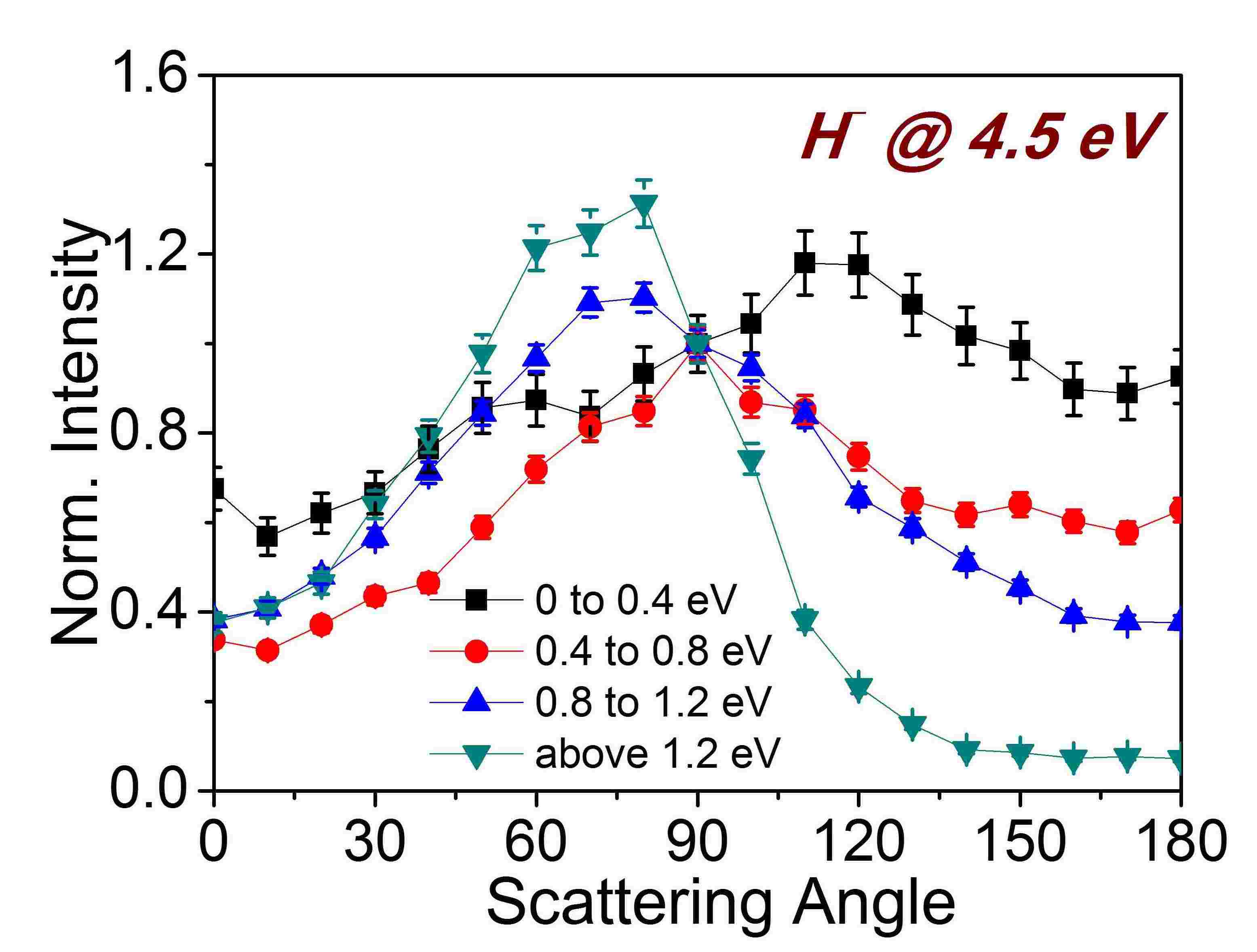}}
\subfloat[]{\includegraphics[width=0.45\columnwidth]{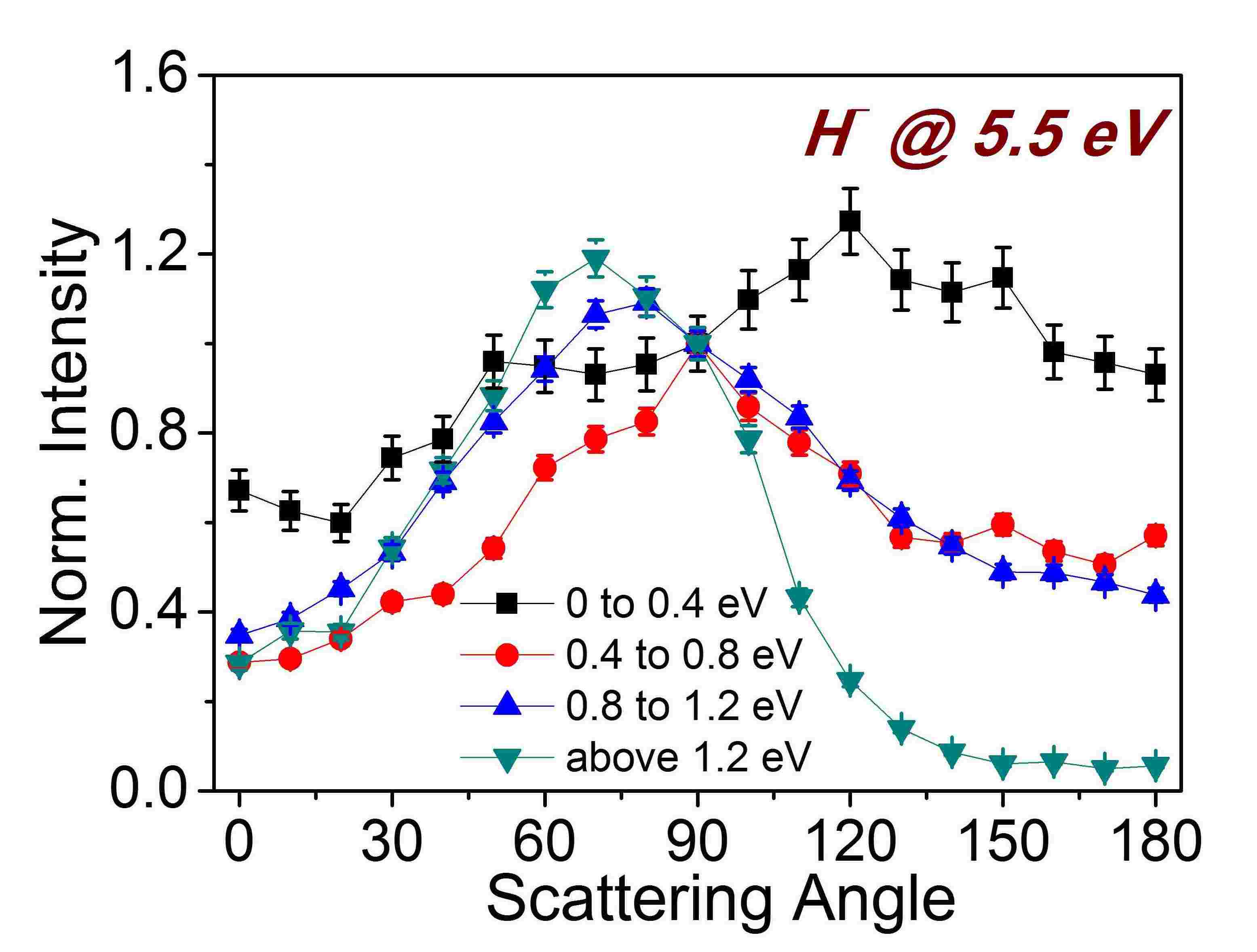}}\\
\subfloat[]{\includegraphics[width=0.45\columnwidth]{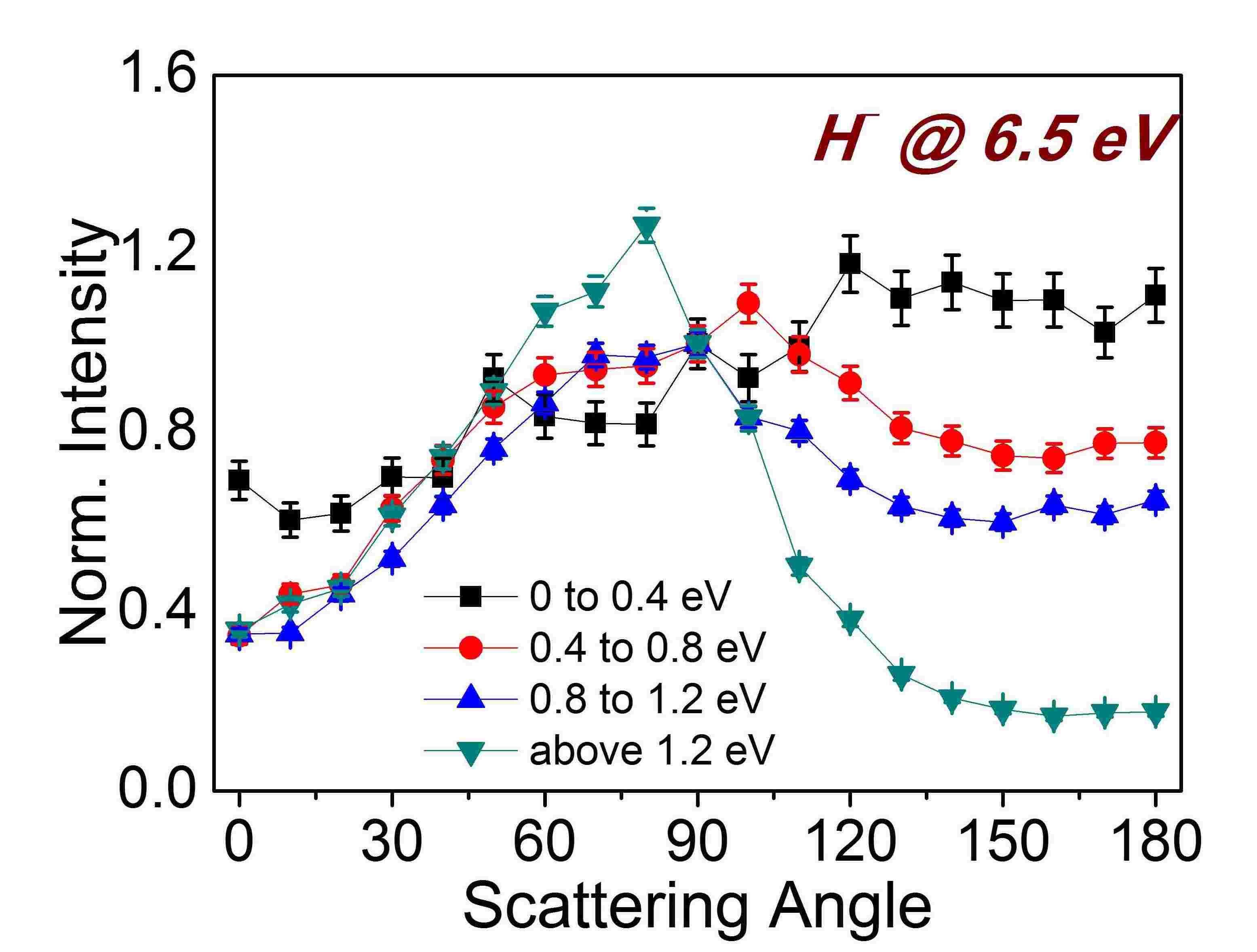}}
\subfloat[]{\includegraphics[width=0.48\columnwidth]{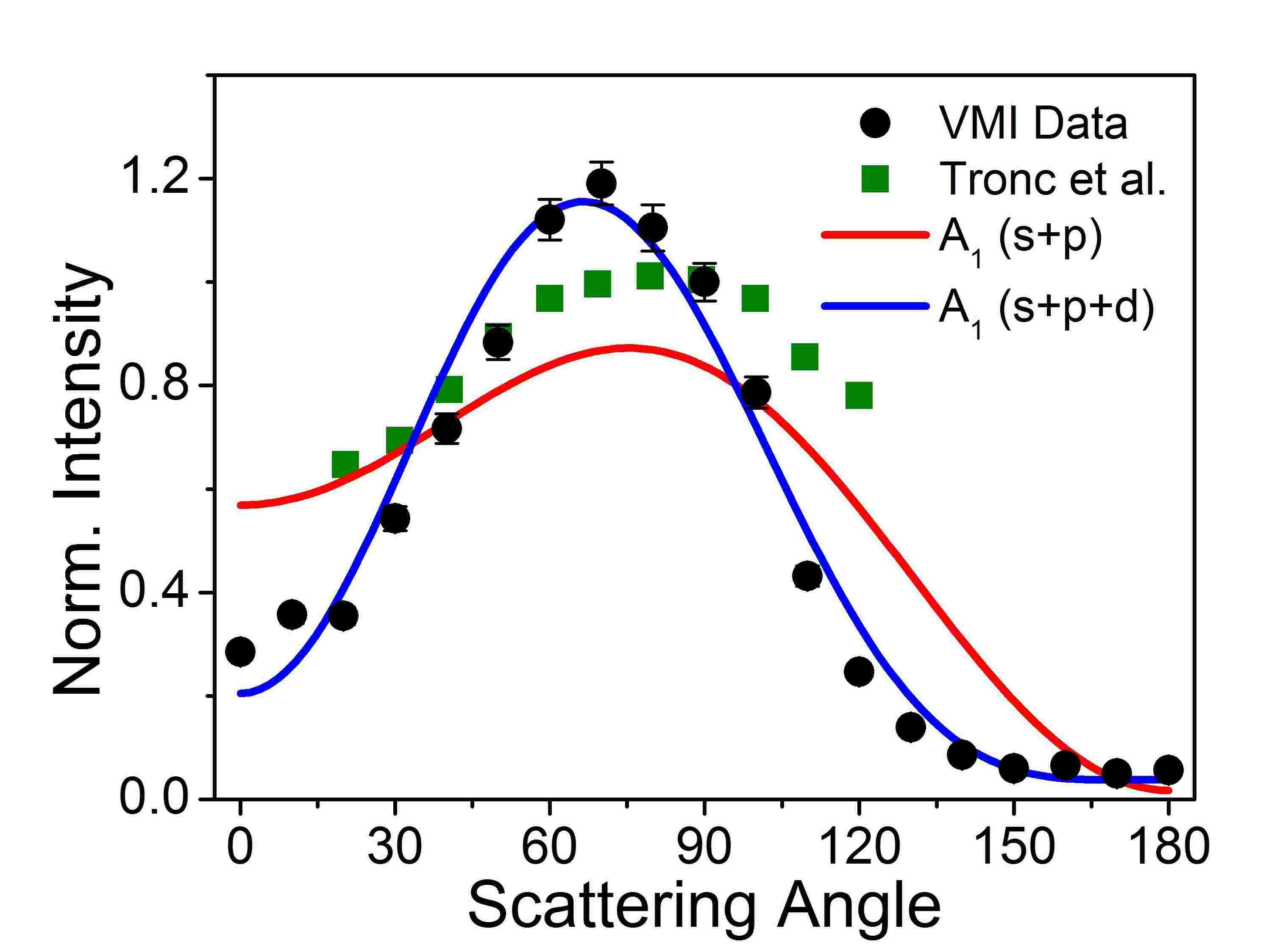}}
\caption{Variation of angular distribution of \ce{H-} ions as a function of the KE is shown at electron energies (a) 4.5 eV (b) 5.5 eV and (d) 6.5 eV. The plots at each electron energy show angular distribution changing as function of KE or internal state of \ce{NH2} fragment due to umbrella mode vibrations of the \ce{NH3^{-*}}. (d) Angular distribution of \ce{H-} ions for 5.5 eV electron energy with maximum kinetic energy (above 1.2 eV) (black circles) and comparison with the measurement of Tronc et al. \cite{c5tronc} (green squares) at 5.7 eV incident electron energy and \ce{H-} ion kinetic energy equal to 1.3 eV. The red and blue solid curves are fits obtained using \ce{A1} symmetry functions taking $s+p$ and $s+p+d$ partial waves respectively.}
\label{fig5.6}
\end{figure}

We have analyzed the angular distribution of \ce{H-} ions as a function of their kinetic energy. The results from three different electron energies are shown in Figure \ref{fig5.6}(a), (b) and (c). For \ce{H-} ions with maximum kinetic energy (i.e. above 1.2 eV), the angular distribution peaks close to $70^{\circ}$ and falls off rapidly at forward and backward scattering angles with an asymmetry i.e. forward angles are more intense than the backward. However, for lower kinetic energies, the angular distribution changes and the backward angles tend to get more intense.  The behaviour of the angular distribution with kinetic energy suggests deviation from axial recoil approximation due to excitation of the umbrella mode vibrations of the \ce{NH3^{-*}} molecular anion. This can be understood in the following way. The ground state equilibrium geometry of \ce{NH3} is a pyramid with N atom at the top and the three H atoms at the base of the pyramid. The three NH bonds are oriented at an angle $68.2^{\circ}$ with respect to the \ce{C3} axis passing through the N atom and the centre of the triangle formed by the three H atoms. Based on the fit for the measured angular distribution (as discussed below) this resonance is formed by excitation of the \ce{3a1} orbital (highest occupied molecular orbital).  This orbital has the electron density distributed above and below the plane perpendicular to this \ce{C3} axis and is zero in the plane. When the electron attachment occurs along the \ce{C3} axis exciting the \ce{3a1} orbital, the energy is coupled to the N-H bonds. The \ce{H-} ions that dissociate instantly carry away almost all the excess energy as translational KE and retain the orientation information of the dissociating bond in the angular distribution. Therefore, for the most energetic ions the angular distribution peaks at $70^{\circ}$ (close to $68.2^{\circ}$) and falls off at other angles in accordance with the ground state equilibrium geometry. However, when the excess energy is used to induce umbrella mode vibrations, this reduces the KE of \ce{H^{-}} and increases the angle between the N-H bonds and the \ce{C3} axis from $70^{\circ}$ and beyond $90^{\circ}$. When the dissociation occurs in this inverted geometry, \ce{H-} ions are scattered in the backward angles and are seen to be intense at lower KE values of \ce{H-}.      

In order to identify the symmetry of the resonance, we use the angular distribution of \ce{H-} ions of kinetic energy greater than 1.2 eV (black circles) as they are expected to be produced under axial recoil approximation.  The result is shown in Figure \ref{fig5.6}(d) and compared with the measurement of Tronc et al. \cite{c5tronc} - the only existing angular distribution measurement on Ammonia till date. Tronc et al. \cite{c5tronc} reported the angular distribution of \ce{H-} ions for 5.7 eV incident electron energy and kinetic energy 1.3 eV (green squares) in the angular range $20^{\circ}$ to $120^{\circ}$.  Comparison with our data shows a broad agreement with a peak about $80^{\circ}$ and intensity falling off at lower and higher angles. However, the intensities in the forward-backward angles are quite different and also the nature of forward-backward asymmetry about the peak. While our data shows higher intensities at forward angles, measurement by Tronc et al. \cite{c5tronc} shows backward angles to be intense. The results by Tronc et al. \cite{c5tronc} appear to be closer to that obtained by us for ions of lower kinetic energy. 

We fit our angular distribution data with \ce{A1} symmetry functions and find that fit involving $s$, $p$ and $d$ waves together matches the distribution very well including the forward-backward asymmetry as compared to the fit with only $s$ and $p$ waves. In the $s+p$ fit (red curve), ratio of $s$ to $p$ is found to be 1:5 with almost no phase difference. The $s+p+d$ fit (blue curve) gives the ratio of the three waves to be 1 : 2.1 : 1.61 with phase differences 0.94 and 0.60 radians between $s$ \& $p$ and $p$ \& $d$ respectively. While these angular fits having been obtained for the molecule in \ce{C_{3v}} geometry, it has been established that the parent state of the 5.5 eV resonance \ce{(3a1)^{-1} (3sa1)^{1}} has a planar configuration resulting in $D_{3h}$ symmetry. Under the $C_{3v}$ symmetry operations, \ce{A1} state is equivalent to the \ce{A1^{'}} and \ce{A2^{"}} states of \ce{D_{3h}} group. Considering that the \ce{3a1} molecular orbital has a dominant $p_{z}$ orbital contribution from the Nitrogen atom, the symmetry of this orbital correlates to the \ce{A2^{"}} state in \ce{D_{3h}} group. The planar symmetry of the parent state of the negative ion state indicates that the equilibrium geometry of the negative ion state also may be planar. Thus the resonance formed in the Franck-Condon region in the \ce{C_{3v}} geometry is expected to relax into its planar configuration on electron attachment, setting in motion the umbrella mode vibrations giving rise to the observed strong deviation from the axial recoil approximation in the angular distribution of \ce{H-} ions of lower kinetic energy.

The conformity with the \ce{A1} symmetry fits implies that the first resonance in Ammonia to be due to the excitation of doubly occupied nitrogen lone pair orbital (\ce{3a1 -> 3sa1}). This excitation is known to change the geometry of the molecule from bent pyramid to a planar one with out-of-plane $n\nu_{2}$ vibrations. Rotational spectroscopy measurements have shown the first excited electronic state of \ce{NH3} (\~{A} \ce{^{1}A2^{"}}) and the lowest cation state to have planar equilibrium geometries (\ce{D_{3h}} symmetry group) \cite{c5walsh}. The out of plane vibrations are also observed in the cation ground state and lowest excited Ryberg state of Ammonia. Potential energy surface of the temporary anion is very similar to that of the Rydberg excited state at least near the initial geometry and the dissociation time of the anion will be close to that for predissociation of Rydberg state. It has also been established that there is no $\nu_{1}$ excitation by comparing data from rotationally cooled molecules with those at room temperature. Further, the \ce{NH3^{-*}} resonance state decay is similar to the lowest excited Rydberg state (\ce{^{1}A2^{"}}) by a predissociation mechanism which arises from excitation of a \ce{3a1} electron to a mixed Rydberg (\ce{3sa1})/antibonding valence ($\sigma^{*}$) orbital. It is generally known that in the case of predissociation, the life time of the excited state is long enough comparable to the molecular rotation period. The effect of rotation may also be contributing to the deviation from the axial recoil approximation observed for lower kinetic energy ions in the angular distribution seen in Figure \ref{fig5.6}(a),(b) and (c), in addition to the bending mode vibrations. 

\begin{figure}[!htbp]
\centering
\includegraphics[width=0.6\columnwidth]{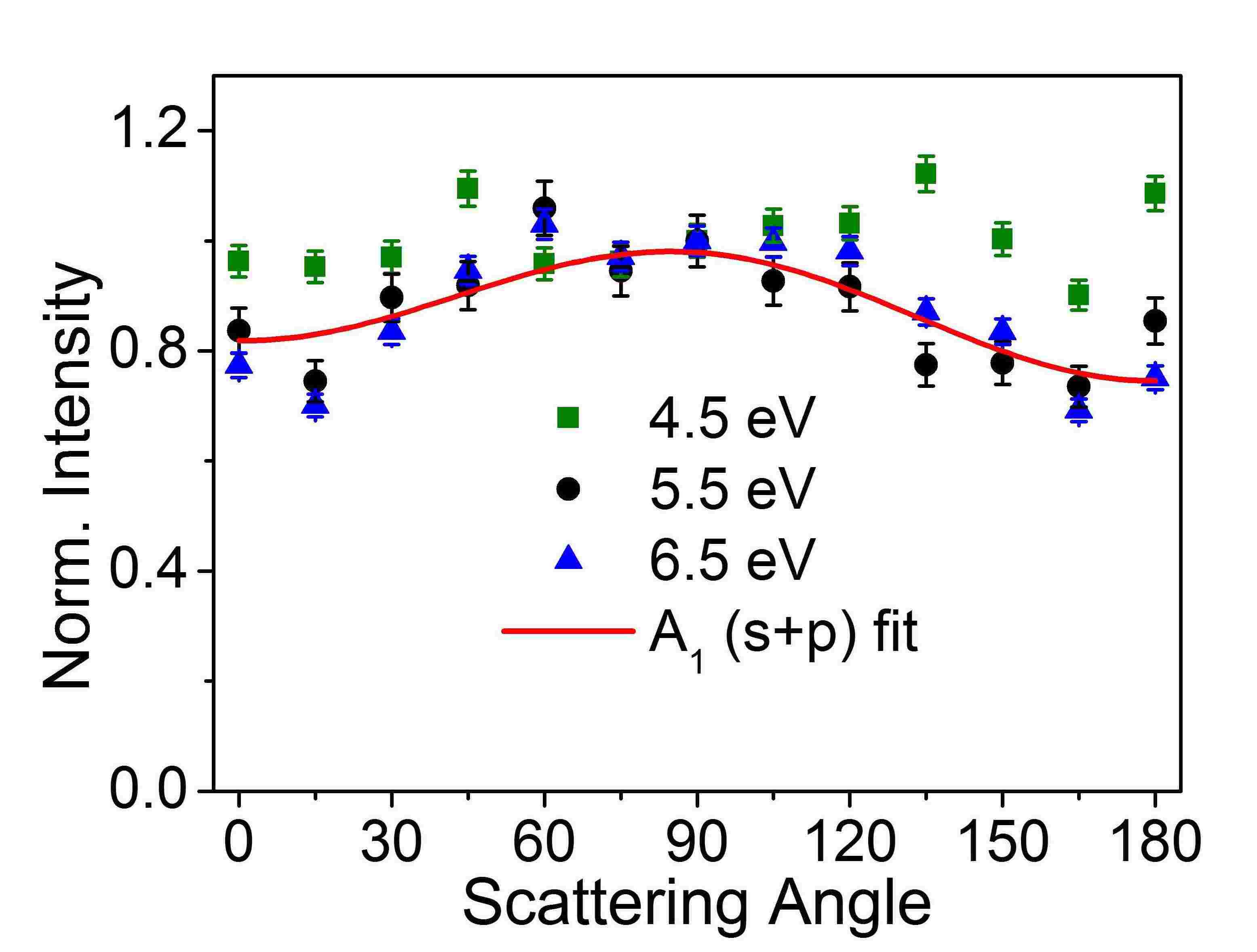}
\caption{Angular distribution of \ce{NH2^{-}} ions at 4.5 eV, 5.5 eV and 6.5 eV. The red curve is the best fit to the 5.5 eV data obtained using \ce{A1} symmetry functions with $s$ and $p$ partial waves respectively. The ratio of the partial waves is found to be $s:p$=1:0.7 with a relative phase difference of 1.52 radians (close to $\pi$/2 radians)}
\label{fig5.7}
\end{figure}

The angular distribution of \ce{NH2^{-}} ions (Figure \ref{fig5.7}) looks more or less isotropic with a broad peak about $90^{\circ}$. It is to be mentioned that the small size of the image doesn't allow for sufficient angular resolution and hence, the isotropy. We obtain a very good fit using \ce{A1} symmetry functions with $s$ and $p$ partial waves. The relative amplitudes are in the ratio $s:p$ = 1:0.7 and $\delta$=1.52 radians (close to $\pi$/2 radians). 

\subsection{Second resonance at 10.5 eV}

So far, there has been no report addressing the symmetry of the second resonance in Ammonia. From the measurements on fragment anions, we find the dissociation channels to be \ce{H- + NH2^{*}(^{2}A1)} and \ce{H + NH2^{-}(^{1}A1)}.  Based on Wigner-Witmer correlation rules, these dissociation channels correlate to either \ce{A1} or E symmetry of the resonance - former coming from a \ce{(3a1)^{-1} (3pe)^{2}} configuration and the latter coming \ce{(1e)^{-1} (3sa1)^{2}} configuration. There are reasons to attribute the second resonance process to excitation of the 1e orbital. The \ce{1e -> 3sa_{1}^{$\prime$}} Rydberg transition in Ammonia is expected to occur at 85000 $cm^{-1}$ i.e. 10.6 eV, whereas the excitation of the lone pair \ce{3a1} to \ce{3pe^{$\prime$}} is seen to occur at 59225 $cm^{-1}$ i.e. 7.4 eV \cite{c5robbins}. Further, the photoelectron spectra of ammonia \cite{c5branton} show two ionization processes corresponding to ionization of the \ce{3a1} and 1e molecular orbitals at 10.15 eV and 14.92 eV respectively. Rydberg transitions converging to these ionization potentials will occur at lower energies and therefore, we expect the electron attachment resonance at 10.5 eV to be due to the excitation of the 1e orbital giving rising to E symmetry of the resonance. Excitation of such a doubly degenerate 1e orbital is expected to have a Jahn-Teller effect.

Figures \ref{fig5.3}(d), (e) and (f) show the velocity map images of \ce{H-} ions at 9.5 eV, 10.5 eV and 11.5 eV.  The angular distribution is strongly in the backward direction. And Figures \ref{fig5.4}(d), (e) and (f) show \ce{NH2^{-}} ions scattered in forward direction. The angular distributions of both the fragment ions are reminiscent of the angular distributions of ions produced from the dissociation of the \ce{^{2}B2} resonance in water at 11.8 eV, where the \ce{H-} ions are scattered backward and the \ce{O-} ions are scattered in the forward direction.

\begin{figure}[!htbp]
\centering
\subfloat[]{\includegraphics[width=0.45\columnwidth]{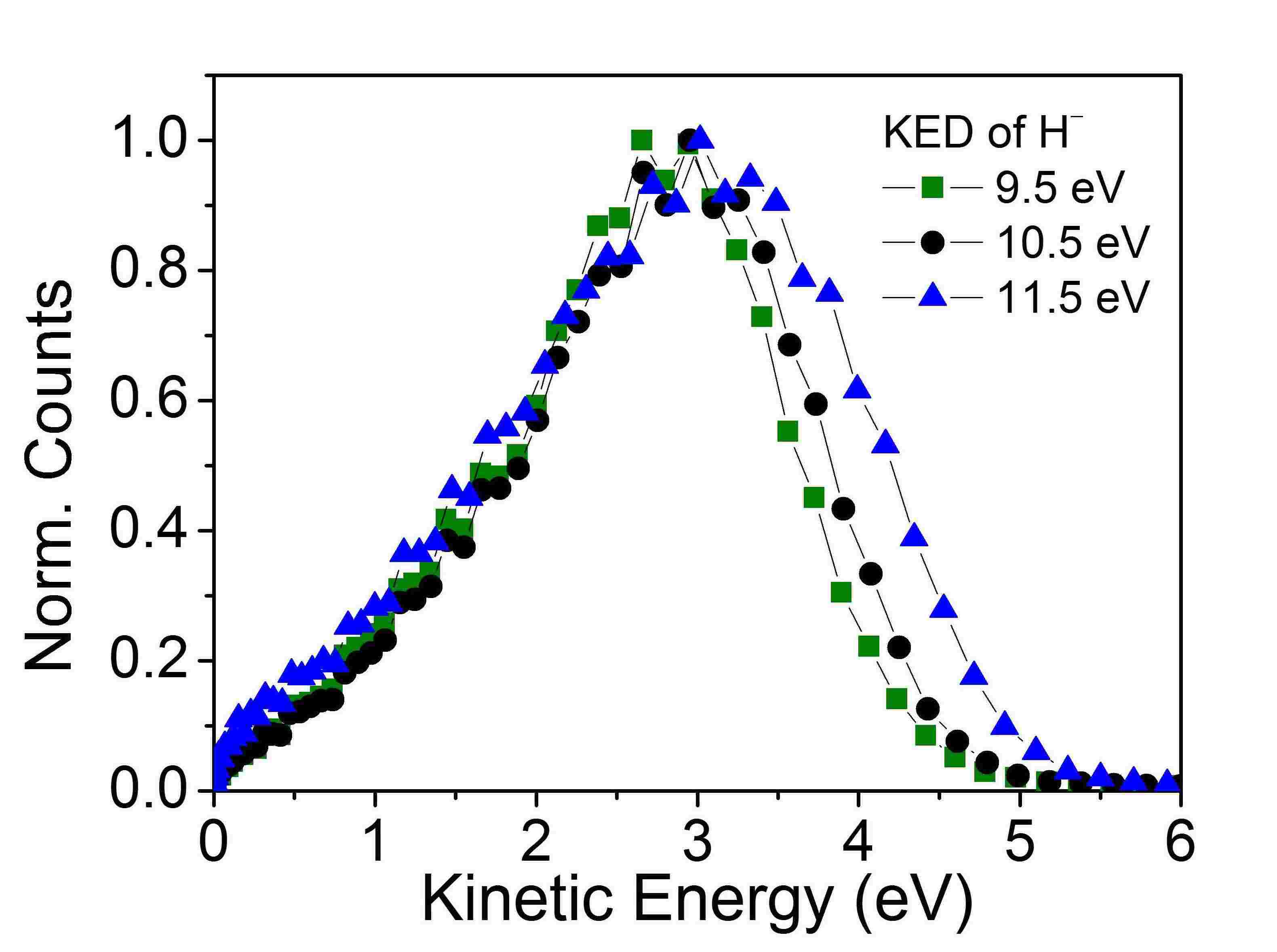}}
\subfloat[]{\includegraphics[width=0.45\columnwidth]{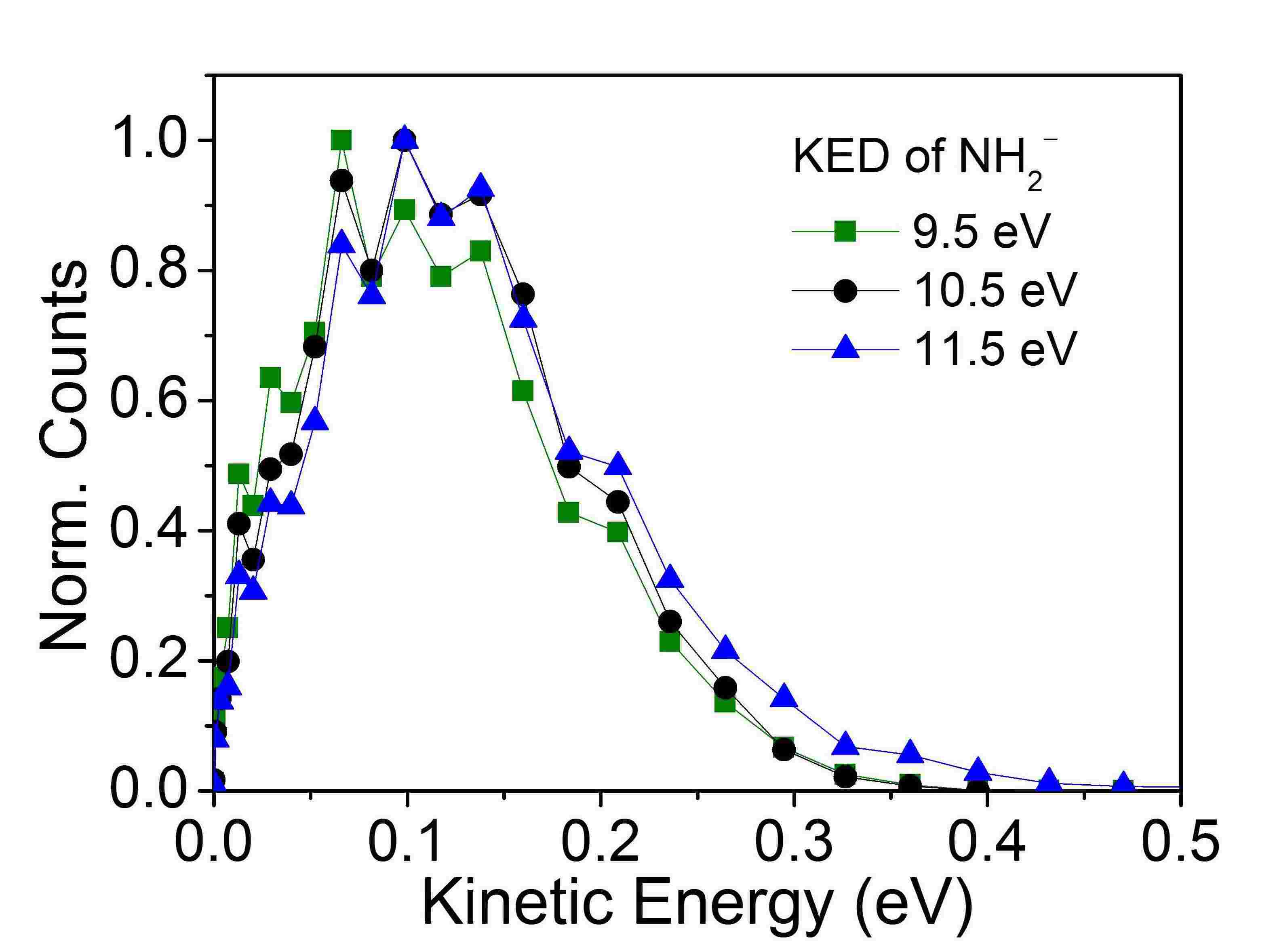}}
\caption{Kinetic energy distribution of (a) \ce{H-} ions and (b) \ce{NH2^{-}} at 9.5, 10.5 and 11.5 eV across the second resonance.}
\label{fig5.8}
\end{figure}

The kinetic energy of \ce{H-} ions ranges from 0 to a maximum of about 5 eV (Figure \ref{fig5.8}(a)) at 10.5 eV electron energy. This suggests the dissociation channel to be \ce{H- + NH2^{*}(^{2}A1)} (threshold: 5.03 eV), where the \ce{NH2} fragment is in an electronically excited state. When the products \ce{H-} and \ce{NH2} are in their ground state (threshold: 3.8 eV), an excess energy of 6.7 eV is available to be shared amongst the two fragments. The maximum KE of \ce{H-} in such a case would be about 6.3 eV, which is higher than what we see. Whereas for \ce{H- + NH2^{*}(^{2}A1)} channel, the maximum KE of \ce{H-} is estimated to be 5.2 eV. This is close to what we observe i.e. 5 eV. Hence, the dissociation channel is found to be \ce{H- + NH2^{*}(^{2}A1)}. 

In the case of \ce{NH2^{-}} ions, we find the maximum kinetic energy to be about 0.35 eV. (see Figure \ref{fig5.8}(b)). Now, reconciling with the possible dissociation channels, it is found that the \ce{H + NH2^{-}(^{1}A1)} channel with threshold of 3.30 eV would have lead to \ce{NH2^{-}} fragment having maximum kinetic energy of 0.42 eV. As we see in Figure \ref{fig5.8}(b), the maximum observed KE of the amino anion is well below the estimated value of 0.42 eV. This could be attributed to the vibrational excitation of \ce{NH2^{-}}. Sharp and Dowell \cite{c5sharpdowell} speculated that the amino anion fragment produced at this resonance is in its first electronically excited state. A rough calculation suggests that presence of \ce{H + NH2^{-*}} channel with appearance energy, say 5 eV would produce \ce{NH2^{-}} ions with energies close to 0.3 eV, close to what we find. However, there are no reports or data available that shed light on the stability / lifetime of the amino anion in the first excited state. In view of the scarcity of information on the excited states of amino anion, we cannot confirm the possibility of amino anion in the first excited state. The formation of vibrationally excited amino anion in ground electronic state may explain the observed KE distribution.  

Alternatively, we also checked for the possibility of amide anion (\ce{NH-}) arising from the three body break up scheme \ce{H + H + NH- (^{2}$\Pi$)} (threshold: 7.94 eV). Wigner-Witmer rules correlate such a three body channel with \ce{NH^{-}(^{2}$\Pi$)} state to E symmetry of the Ammonia anion resonance.  At an incidence energy of 10.5 eV, the excess energy available would be 2.56 eV and assuming this is distributed amongst the three fragments as translational kinetic energy, the maximum KE of \ce{NH-} would be is 2/17th of 2.56 eV i.e. close to 0.3 eV. Assuming the two hydrogen atoms to break symmetrically from the \ce{NH-}, the KE of \ce{NH-} could be expressed in terms of half the bond angle ($\theta$) between H-NH-H i.e. $E_{\ce{NH-}}=2E_{o}\cos^{2}\theta/(15+2\cos^{2}\theta)$ where $E_{o}$=2.56 eV. We observe that with $\theta$ varying from $0^{\circ}$ to $90^{\circ}$, the KE of \ce{NH-} decreases from 0.3 to 0 eV. The peak value of 0.12 eV corresponds to $\theta=53^{\circ}$ (i.e. H-NH-H bond angle is$106^{\circ}$). Thus, symmetry and energy arguments allow for the three body channel giving rise to \ce{NH-} to exist. However, measurements by Sharp and Dowell \cite{c5sharpdowell} have ruled out the presence of \ce{NH-} channel. 

Thus, there are three candidates (\ce{NH2^{-}}, \ce{NH2^{-*}} and \ce{NH-}) that could explain the observed KE distribution of the heavier anion fragment at the second resonance process. Based on our data and the available literature, we conclude that the observed heavier anion is \ce{NH2^{-}} in ground electronic state with internal excitation. It is highly desirable that potential energy surface calculations of \ce{NH3^{-*}} resonance state and experiments with improved electron energy and mass resolution substantiate or challenge this conclusion.

\begin{figure}[!htbp]
\centering
\subfloat[]{\includegraphics[width=0.32\columnwidth]{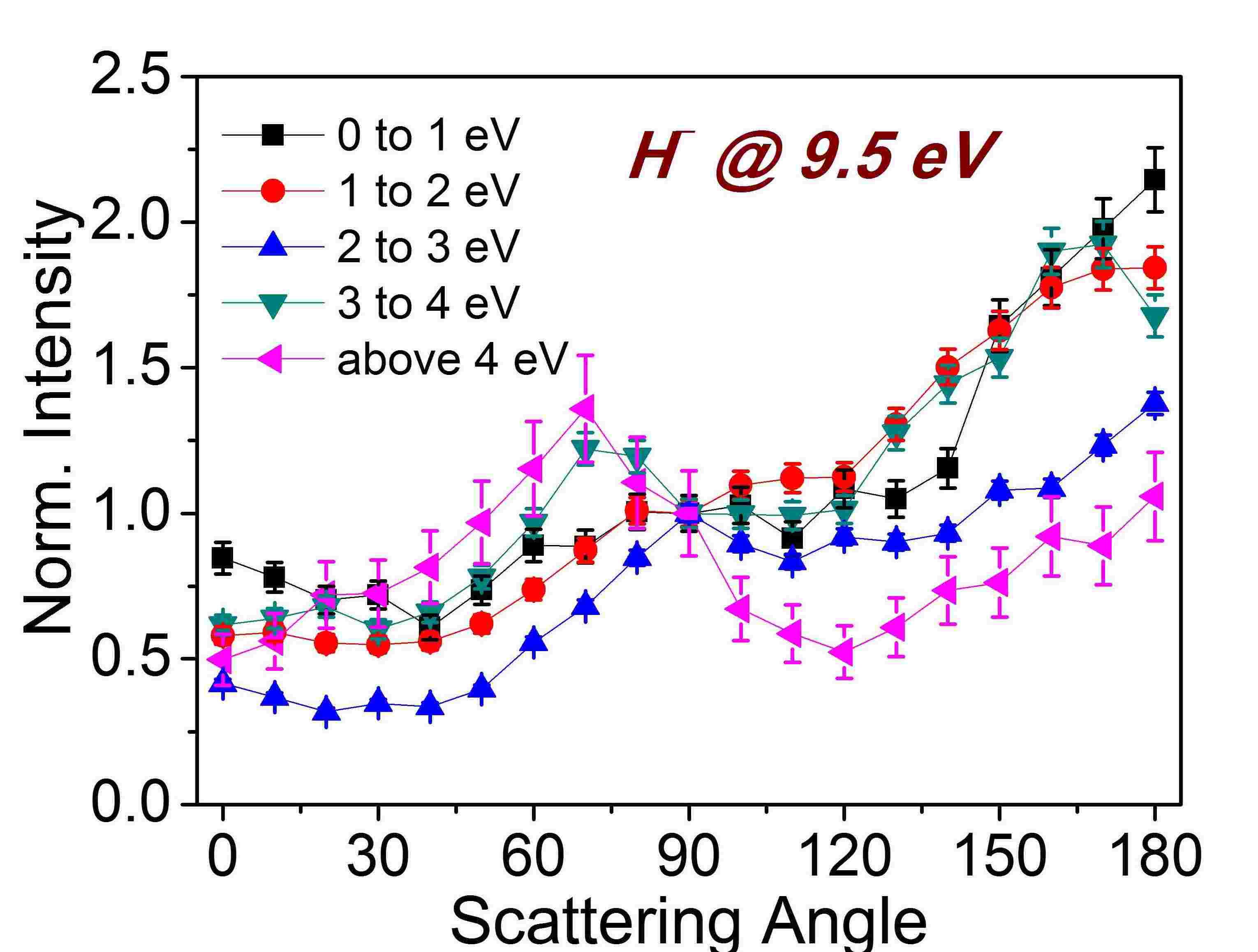}}
\subfloat[]{\includegraphics[width=0.32\columnwidth]{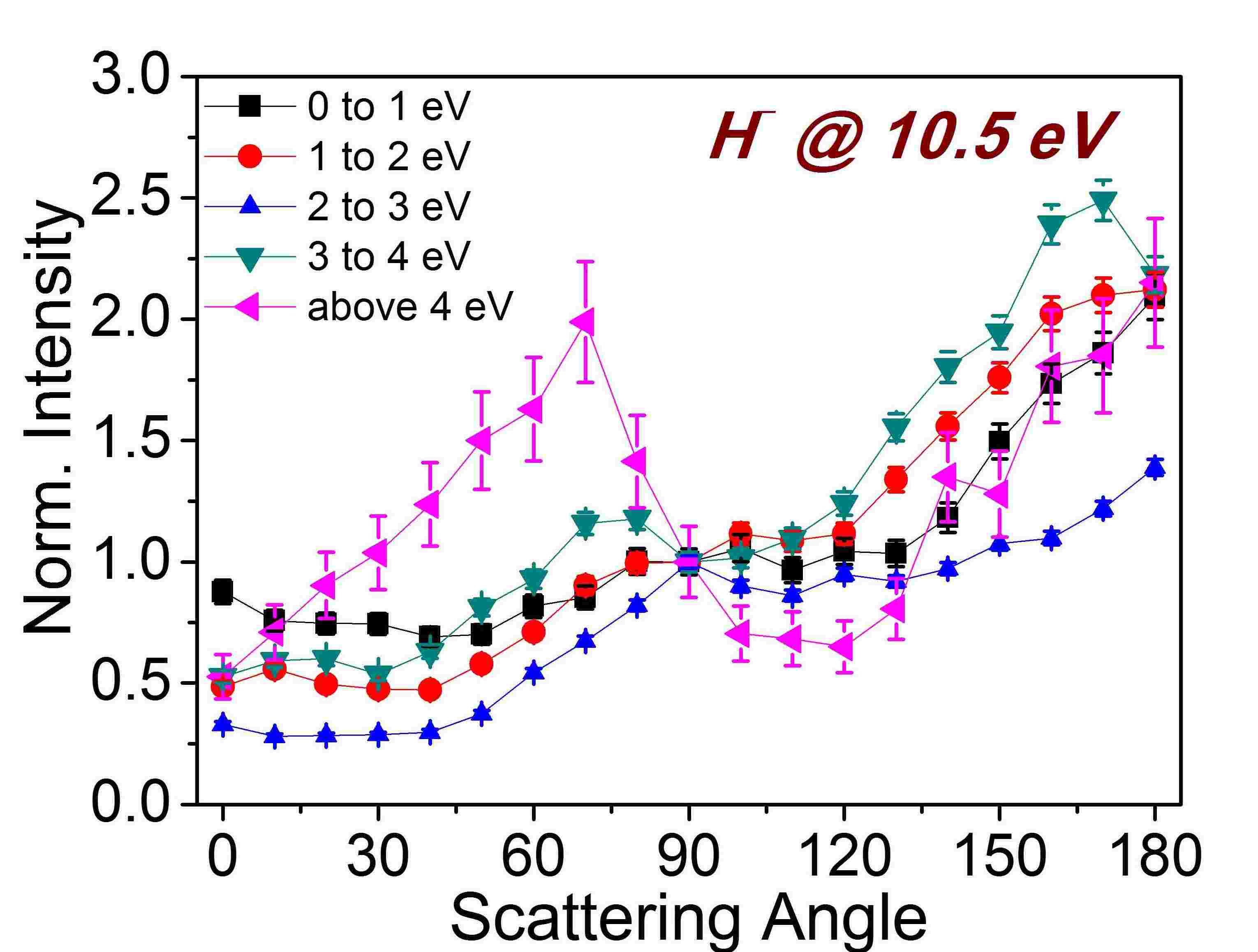}}
\subfloat[]{\includegraphics[width=0.32\columnwidth]{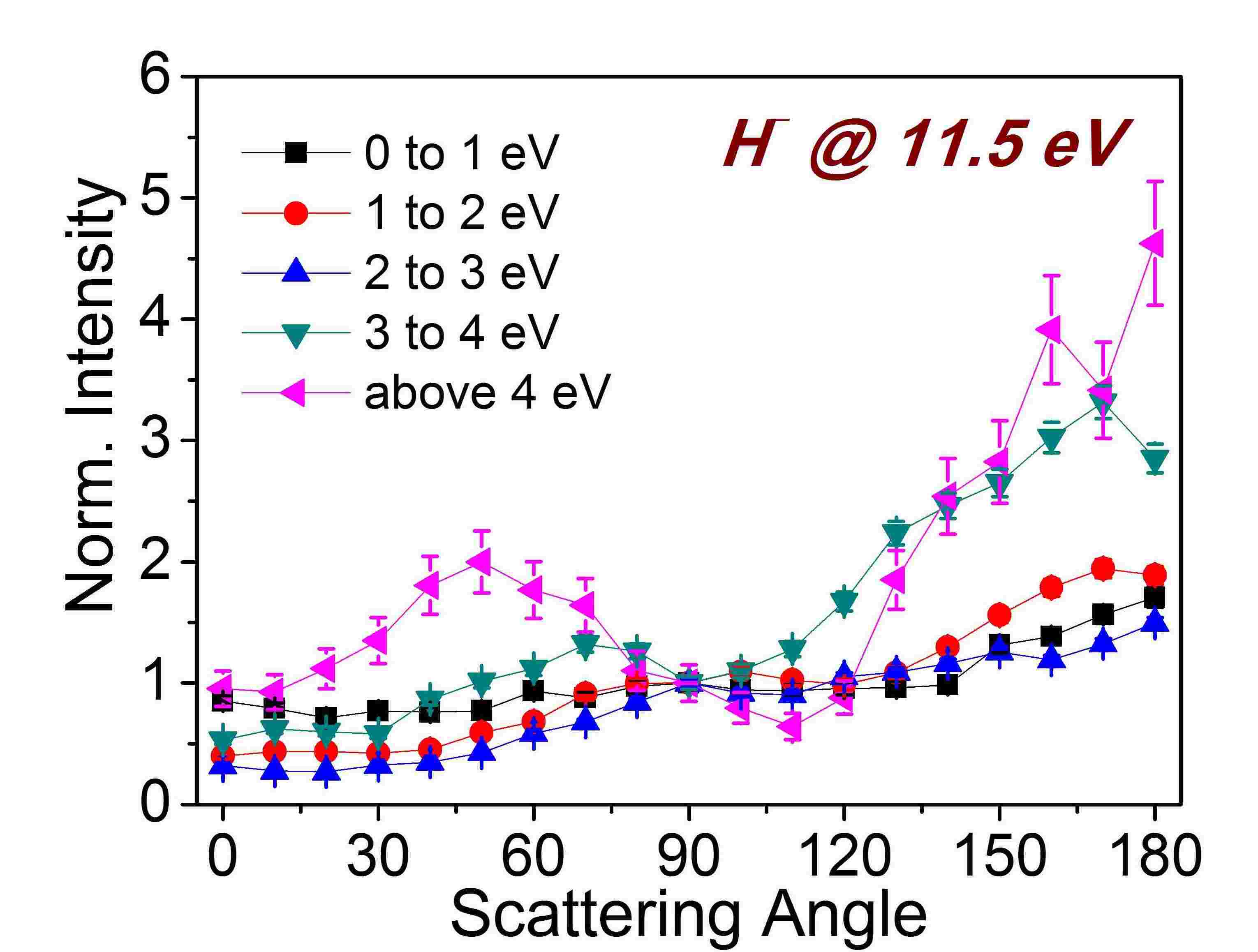}}\\
\subfloat[]{\includegraphics[width=0.32\columnwidth]{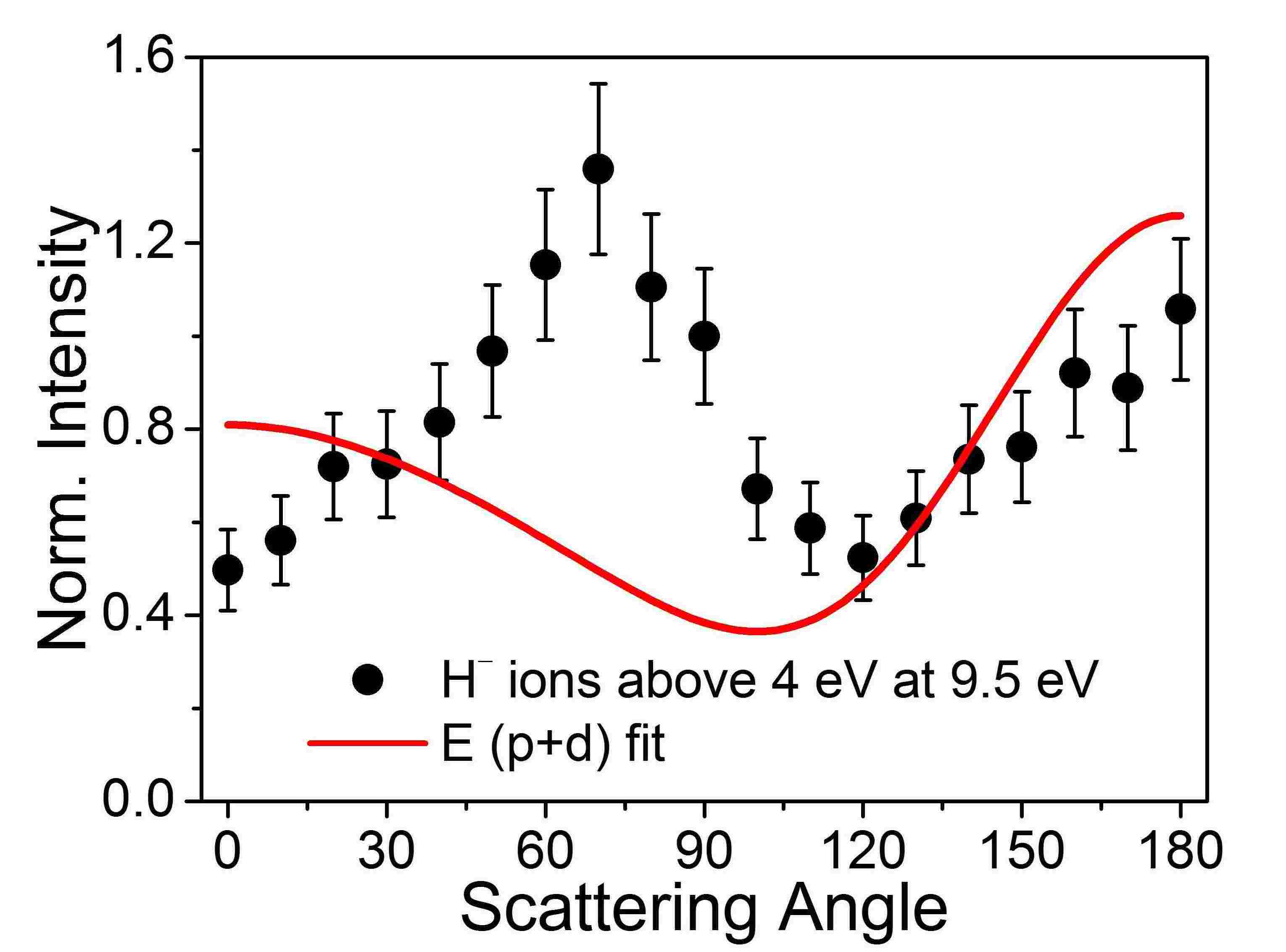}}
\subfloat[]{\includegraphics[width=0.32\columnwidth]{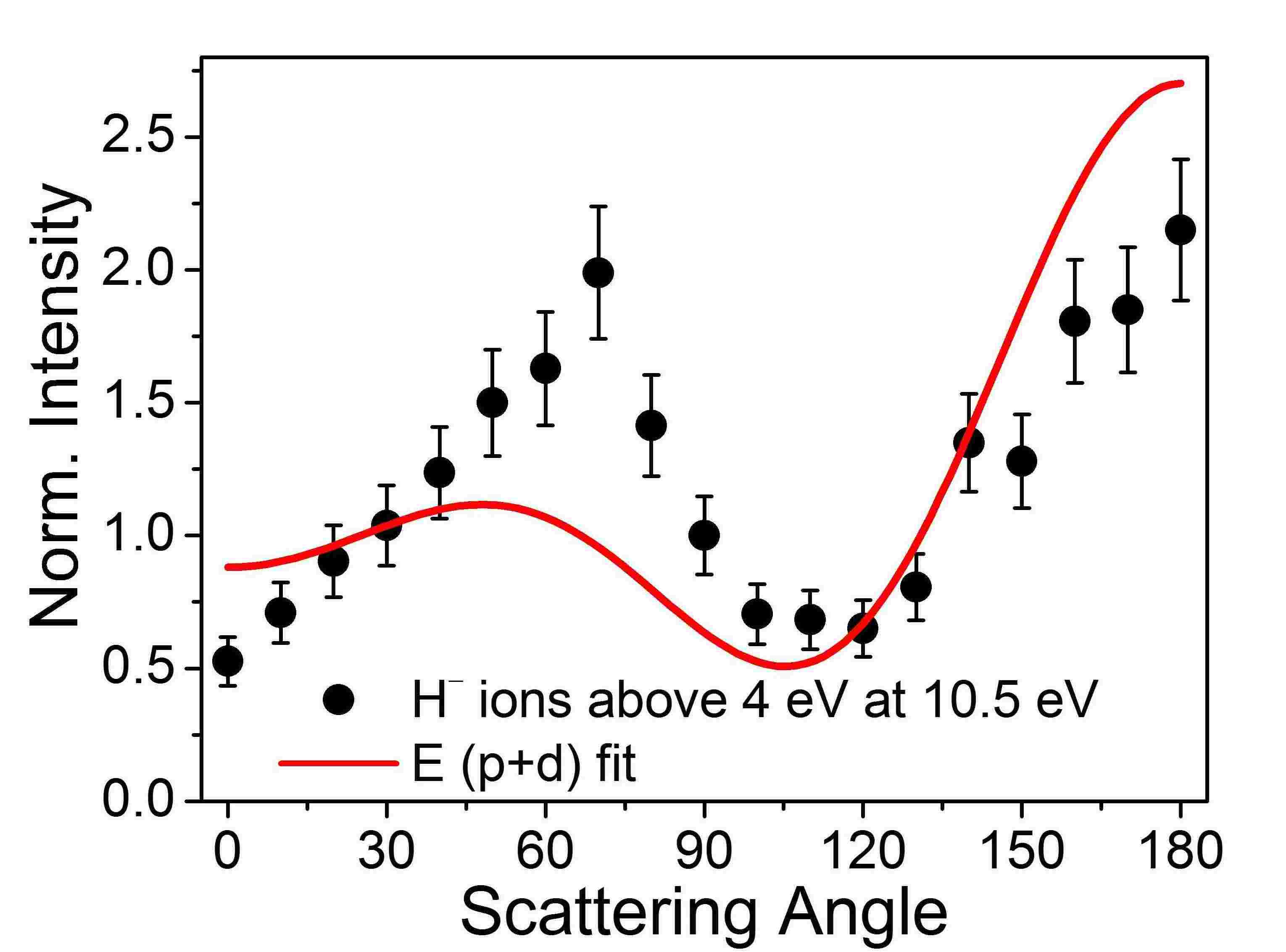}}
\subfloat[]{\includegraphics[width=0.32\columnwidth]{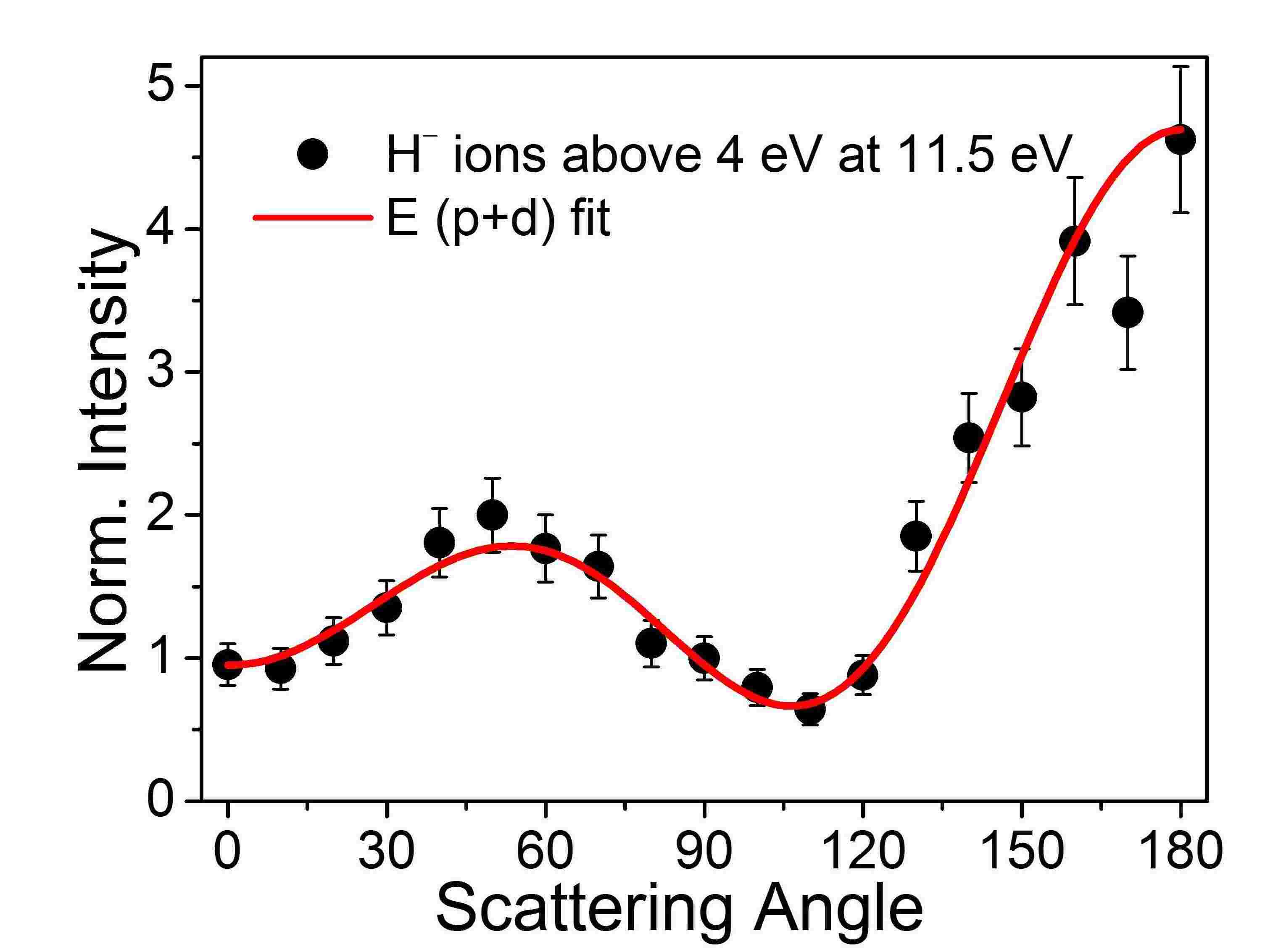}}
\caption{Angular distribution of \ce{H-} ions as a function of its kinetic energy at incident electron energies (a) 9.5 eV (b) 10.5 eV and (c) 11.5 eV. The variation in angular distribution shows rearrangement of the molecular geometry prior to dissociation due to vibrational motion suggesting deviation from axial recoil approximation. To  identify the symmetry of the resonance, we fit the angular distribution data of \ce{H-} ions above 4 eV at electron energies (d) 9.5 eV, (e) 10.5 eV and (f) 11.5 eV assuming E symmetry. The fits fairly agree with the observed data especially at 10.5 eV and 11.5 eV}
\label{fig5.9}
\end{figure}

\begin{figure}[!htbp]
\centering
\includegraphics[width=0.5\columnwidth]{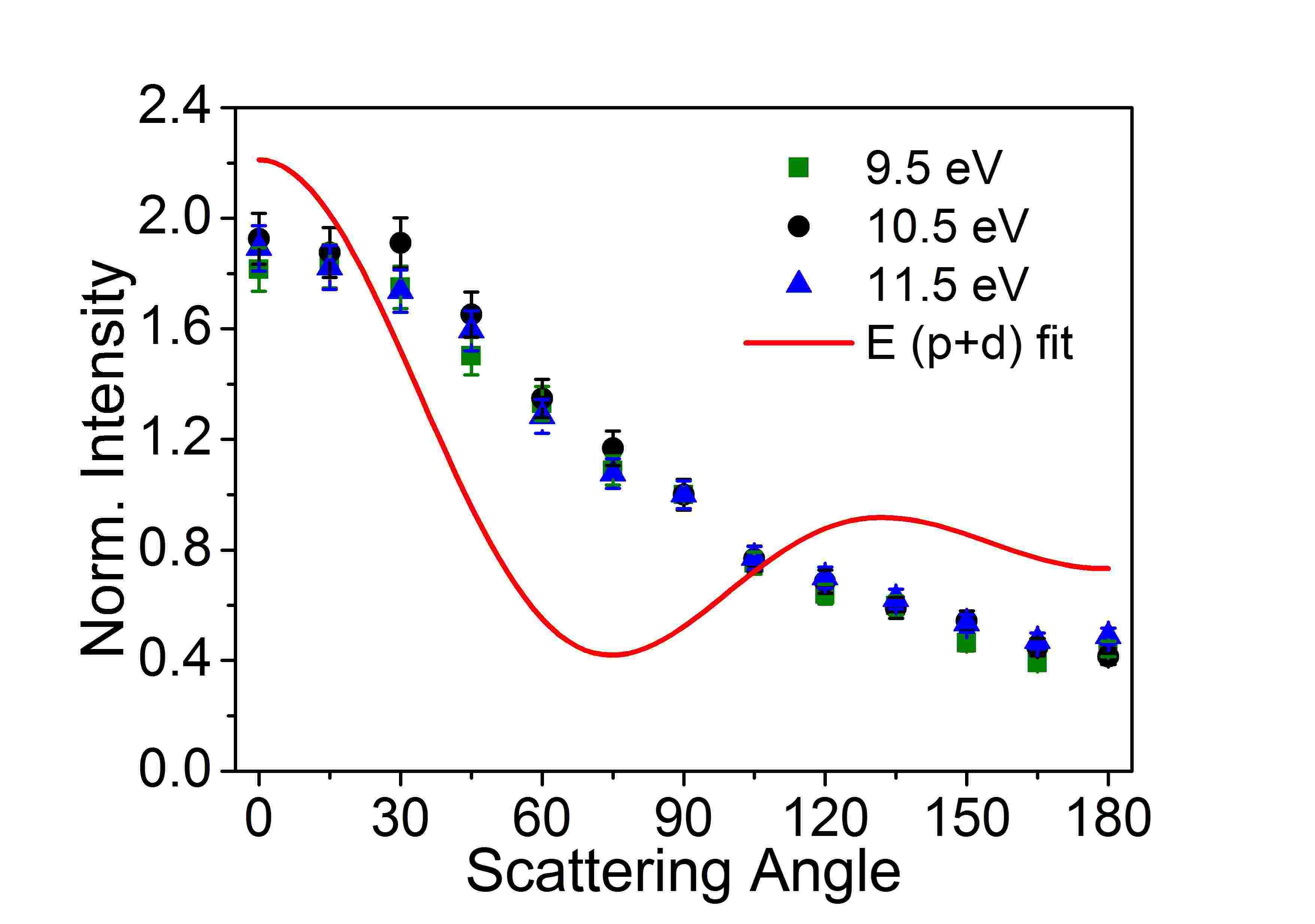}
\caption{Angular distribution of \ce{NH2^{-}} ions is shown across the resonance at energies 9.5, 10.5 and 11.5 eV. The red curve is the fit for two body breakup using E symmetry functions taking $p$ and $d$ partial waves in the ratio 1:5.8 with phase difference of 0 radians. The fit qualitatively reproduces the forward - backward asymmetry.}
\label{fig5.10}
\end{figure}

Analyzing the angular distribution of \ce{H-} ions as a function of the kinetic energy release we see variation in the angular distributions indicating structural changes of the ammonia anion. Figures 9(a), (b) and (c) show the angular plots of \ce{H-} ions as function of KE. It is seen that the ions with maximum KE (i.e. above 4 eV) show an angular distribution with peaks at $60^{\circ}$ and $180^{\circ}$. As the kinetic energy decreases, the backward angles become more intense. We explain the observed angular distribution and its variation with the kinetic energy of the ion in terms of the electron attachment to the doubly degenerate 1e orbital. The electron density in the 1$e$ orbital is seen to be distributed closer to the plane of the molecule. When the incoming electron gets captured in the plane of the molecule along one of the H-N bonds oriented along the electron beam, instant dissociation of the ammonia anion may lead to the breaking any of the three N-H bonds in this orientation. The ions with largest kinetic energy correspond to the instantaneous fragmentation before the excess energy could be redistributed into other vibrational modes. While one N-H bond is antiparallel to the electron beam, the other two are oriented at $60^{\circ}$ approximately on either sides of the electron beam. Thus, we see peaks at $60^{\circ}$ and $180^{\circ}$ in the angular distribution of \ce{H-} ions with highest kinetic energy (Figures \ref{fig5.9}(a), (b) and (c)) and the fits in (d), (e) and (f) fairly match the observed angular distribution in accordance with the axial recoil approximation. However, the excess energy in the system can go into the excitation of the vibrational modes and the bending mode may reduce the H-N-H bond angle from $120^{\circ}$ to $90^{\circ}$ while causing the H-N bond along the electron beam to stretch and eventually eject the \ce{H-}. The probability of \ce{H-} ejected from the other two N-H bonds (oriented at close to $90^{\circ}$ with respect to the electron due to bending mode vibrations) seems to be lower than at $180^{\circ}$ direction, giving rise to the backward distribution.

Figure \ref{fig5.10} shows the angular distribution of \ce{NH2^{-}} fragments along with the fit of E symmetry. We fit the observed angular distributions with E symmetry functions taking $p$ and $d$ partial waves. We see that the fit qualitatively reproduces the finite intensities at the forward backward angles along with the asymmetry, but is not a very good fit. The lack of a good fit indicates that the dissociation into \ce{H + NH2^{-}} does not follow axial recoil approximation.

\section{Summary} 
	\begin{enumerate}
		\item Imaging of \ce{H-} and \ce{NH2^{-}} fragment anions across the two resonances peaking at 5.5 eV and 10.5 eV.
		\item	First resonance peaking at 5.5 eV - \ce{A2^{"}} (\ce{D_{3h}}) or \ce{A1} (\ce{C_{3v}}) symmetry.
			\begin{enumerate}
				\item	\ce{H- + NH2 (^{2}B1)} and \ce{H + NH2^{-}(^{1}A1)} dissociation channels confirmed.
				\item	Variation of \ce{H-} angular distribution with kinetic energy - Umbrella mode vibrations ($\nu_{2}$) of Ammonia anion inferred.
			\end{enumerate}

		\item	Second resonance peaking at 10.5 eV - E (\ce{C_{3v}}) symmetry
			\begin{enumerate}	
				\item First measurement of angular distribution.
				\item \ce{H-} and \ce{NH2^{-}} angular distributions strongly backward and forward respectively -	similar to the 11.8 eV resonance in water.
				\item \ce{H-} channel seen to be \ce{H- + NH2^{*}}(threshold: 3.30 eV) where the amino fragment is in first excited state.
				\item Effect of internal excitations of Ammonia anion seen in \ce{H-} angular distribution. 
				\item The heavier anion fragment channel is inferred to be \ce{H + NH2^{-}} where the \ce{NH2^{-}} is in ground electronic state but vibrationally excited.
				\item Symmetry and energy arguments allow for the existence \ce{H + H + NH^{-}(^{2}$\Pi$)} channel, however previous measurements [4] showed no presence.
			\end{enumerate}
		\end{enumerate}

\appendix
\newpage
\section{Angular distribution curves for $C_{3v}$ point group}
Expressions for angular distribution for various symmetries under $C_{3v}$ group taking lowest allowed partial waves.

\section*{$A_{1}$ to $A_{1}$ transition}
\begin{eqnarray}
I_{s}(\theta)&=& 1 \\
I_{p}(\theta)&=&\sin^{2}\beta \sin^{2}\theta + 2 \cos^{2}\beta \cos^{2}\theta \\
I_{d}(\theta)&=& \frac{9}{16} (\sin^{4}\beta \sin^{4}\theta + \sin^{2}2\beta \sin^{2}2\theta)+ \frac{1}{8}(3 \cos^{2}\beta - 1)^{2}(3 \cos^{2}\theta - 1)^{2} \\
I_{s+p}(\theta)&=& a^{2} + b^{2}(\sin^{2}\beta \sin^{2}\theta + 2 \cos^{2}\beta \cos^{2}\theta) + 4ab \cos\beta \cos\theta \cos\delta \\
I_{s+p+d}(\theta)&=& a^{2} + b^{2}(\sin^{2}\beta \sin^{2}\theta + 2 \cos^{2}\beta \cos^{2}\theta) \nonumber \\
&& + c^{2} (\frac{9}{16} (\sin^{4}\beta \sin^{4}\theta + \sin^{2}2\beta \sin^{2}2\theta)+ \frac{1}{8}(3 \cos^{2}\beta - 1)^{2}(3 \cos^{2}\theta - 1)^{2}) \nonumber \\
&& + 4 a b \cos\beta \cos\theta \cos\delta_{1} \nonumber \\
&& + 2 b c (\frac{3}{4}\sin\beta \sin2\beta \sin\theta \sin2\theta + \frac{1}{2}\cos\beta (3\cos^{2}\beta-1)\cos\theta (3\cos^{2}\theta-1)) \cos\delta_{2} \nonumber \\
&& + a c (3\cos^{2}\beta-1)(3\cos^{2}\theta-1) \cos(\delta_{1}+\delta_{2}) 
\end{eqnarray}

\section*{$A_{1}$ to $A_{2}$ transition}
\begin{eqnarray}
I_{f}(\theta)&=& \frac{5}{4}(\cos^{2}\beta + \frac{\sin^{2}\beta}{4})^{2} \sin^{6}\theta + \frac{45}{16} \sin^{2}{2\beta} \sin^{4}\theta \cos^{2}\theta \nonumber \\
&& + \frac{45}{4} \sin^{4}\beta \sin^{2}\theta (\cos^{2}\theta-\frac{\sin^{2}\theta}{4})^{4}
\end{eqnarray}

\section*{$A_{1}$ to $E$ transition}
\begin{eqnarray}
I_{p}(\theta)&=& 2 (\cos^{2}\beta \sin^{2}\theta + 2 \sin^{2}\beta \cos^{2} \theta) \\
I_{d}(\theta)&=&1.5(\frac{1}{4} \sin^{2}2\beta \sin^{4}\theta + \cos^{2}2\beta \sin^{2}2\theta + \frac{1}{2}\sin^{2}2\beta (3\cos^{2}-1)^{2}) \\
I_{p+d}(\theta)&=& 2a^{2} (\cos^{2}\beta \sin^{2}\theta + 2 \sin^{2}\beta \cos^{2} \theta) \nonumber \\
&& + 1.5b^{2} (\frac{1}{4} \sin^{2}2\beta \sin^{4}\theta + \cos^{2}2\beta \sin^{2}2\theta + \frac{1}{2}\sin^{2}2\beta (3\cos^{2}-1)^{2}) \nonumber \\
&& + 2ab \sqrt{3} (\cos\beta \cos2\beta \sin\theta \sin2\theta + \sin\beta \sin2\beta \cos\theta (3\cos^{2}\theta-1)) \cos\delta
\end{eqnarray}

\end{document}